# Octupole-driven spin torque switching of antiferromagnetic tunnel junctions

Jaimin Kang[1*†], Mohammad Hamdi[1*†], Shun Kong Cheung[1], Lin-Ding Yuan[2], Mohamed Elekhtiar[3], Eric Matt[1,4], William Rogers[2,4], Andrea Meo[5], Peter G. Lim[2,4], M.S. Nicholas Tey[2], Anthony D'Addario[6], Shiva T. Konakanchi[7], Jordan Athas[1], Sevdenur Arpaci[1,4], Antoine Devincenti[2], Lei Wan[8], Sanjay C. Mehta[8], Nicholas Zogbi[9,10], Vinod K. Sangwan[2], Jean Anne C. Incorvia[9,10], Jennifer Fowlie[2,4], Pramey Upadhyaya[11], Mario Carpentieri[5], Vinayak P. Dravid[2,4], Mark C. Hersam[1,2,4,12], Jordan A. Katine[8], Gregory D. Fuchs[6], Giovanni Finocchio[13], Evgeny Y. Tsymbal[3], James M. Rondinelli[2,4], Pedram Khalili Amiri[1,4*]

*[1]Department of Electrical and Computer Engineering, Northwestern University, Evanston, Illinois 60208, USA*

*[2]Department of Materials Science and Engineering, Northwestern University, Evanston, Illinois 60208, USA*

*[3]Department of Physics and Astronomy, University of Nebraska, Lincoln, Nebraska 68588, USA*

*[4]Applied Physics Program, Northwestern University, Evanston, Illinois 60208, USA*

*[5]Department of Electrical and Information Engineering, Politecnico di Bari, Bari 70125, Italy*

*[6]School of Applied and Engineering Physics, Cornell University, Ithaca, New York 14853, USA*

*[7]Department of Physics and Astronomy, Purdue University, West Lafayette, Indiana 47907, USA*

*[8]Western Digital Corporation, San Jose, California 95119, USA*

*[9]Chandra Family Department of Electrical and Computer Engineering, The University of Texas at Austin, Austin, Texas 78712, USA*

*[10]Microelectronics Research Center, The University of Texas at Austin, Austin, Texas 78758, USA*

*[11]Elmore Family School of Electrical and Computer Engineering, Purdue University, West Lafayette, Indiana 47907, USA*

*[12]Department of Chemistry, Northwestern University, Evanston, Illinois 60208, USA*

*[13]Department of Mathematical and Computer Sciences, Physical Sciences and Earth Sciences, University of Messina, Messina 98166, Italy*

*† These authors contributed equally to this work.*

** Corresponding authors:*

*jaimin.kang@northwestern.edu, mohammad.hamdi@northwestern.edu, pedram@northwestern.edu*

Magnetic tunnel junctions (MTJs) based on ferromagnets are canonical devices in spintronics, with wide-ranging applications in data storage, computing, and sensing. They simultaneously exhibit mechanisms for electrical detection and control of magnetic order through the tunneling magnetoresistance (TMR)[1-5] and spin-transfer torque (STT)[6-9] effects, respectively. It was long assumed that neither of these effects could be sizeable in all-antiferromagnetic tunnel junctions (AATJs), since they exhibit no net magnetization. Recently, however, it was shown that AATJs based on chiral antiferromagnets do exhibit TMR due to their non-relativistic momentum-dependent spin polarization and cluster magnetic octupole moment (CMO), which are manifestations of their spin-split band structure[10-13]. However, the reciprocal effect, i.e., the antiferromagnetic counterpart of STT, has been assumed non-existent due to the total electric current being spin-neutral. Here, we report nanoscale AATJs exhibiting this reciprocal effect, which we term octupole-driven spin-transfer torque (OTT). We demonstrate current-induced OTT switching of $PtMn_3$|MgO|$PtMn_3$ AATJs, exhibiting a TMR value of 363% at room temperature and switching current densities of the order of 10 MA/cm$^2$. Our theoretical modeling explains the origin of OTT in terms of the imbalance between intra- and inter-sublattice spin currents across the AATJ, and equivalently, in terms of the non-zero net cluster octupole polarization of each $PtMn_3$ layer. This work establishes a new materials platform for antiferromagnetic spintronics and provides a pathway towards deeply scaled magnetic memory and room-temperature terahertz technologies.

## Introduction

A conventional magnetic tunnel junction (MTJ) consists of two ferromagnetic layers sandwiching a thin non-magnetic insulator. Under a vertical electric field, electrons tunneling through the insulator are spin-polarized by the ferromagnetic electrodes due to their spin-split electronic band structure. This results in the tunnelling magnetoresistance (TMR) effect, where electrical resistance of the MTJ depends on the relative orientation between the magnetization of the ferromagnets[1-5]. Reciprocally, the spin-polarized current also exerts torque on the magnetization of a ferromagnetic electrode, leading to spin-transfer torque (STT) switching[6-9]. These two mechanisms serve as reading and writing protocols for commercialized magnetic random-access memory (MRAM) and are the basis for other devices such as spin torque oscillators, spin diodes, and sensors[14,15].

Antiferromagnetic materials have emerged as promising candidates for spintronics due to their potential for higher-density integration, immunity to external fields, and potential ultrafast operation owing to their exchange-dominated dynamics[16-25]. A key goal is to use these materials as active elements of the MTJ structure, replacing the role of ferromagnets while retaining the same read and write functions using tunneling currents through the device[26]. Prominent candidates to fulfill this aim are altermagnets[27-32] and noncollinear antiferromagnets[23,33-36], both breaking the effective time-reversal symmetry (TRS) by specific combinations of their crystal lattice and magnetic structure. Consequently, they exhibit nontrivial electronic transport behaviors due to their spin-split band structure, multipole polarization, or non-vanishing Berry curvature. These characteristics have enabled the observation of various phenomena traditionally associated with ferromagnets, including the anomalous Hall, anomalous Nernst, and magneto-optical Kerr effects, which were extensively investigated in structures with a single antiferromagnetic layer[37-40].

Recently, all-antiferromagnetic tunnel junctions (AATJs) were demonstrated to exhibit a sizable room-temperature TMR by employing noncollinear antiferromagnets such as $PtMn_3$ and $SnMn_3$ for both electrodes[11-13]. Despite their vanishing net magnetization, these materials exhibit a non-relativistic momentum-dependent spin polarization which reverses with opposite spin configurations, leading to finite TMR[41]. This effect can also be viewed as a consequence of polarization of the CMO – an order parameter that describes the noncollinear magnetic structure – with the relative orientation between the octupole moments of the two noncollinear antiferromagnetic electrodes determining the TMR. Switching of the antiferromagnetic order in these experiments was performed either by magnetic fields[11,12] or by spin-orbit torque (SOT)[13] from a heavy metal layer.

Onsager reciprocity suggests that TMR in AATJs should be accompanied by a reciprocal effect, analogous to STT in ferromagnet-based MTJs[42-45]. Building on the octupole-induced mechanism of TMR in these structures, one can hypothesize the existence of a reciprocal torque, which can be termed octupole-driven spin-transfer torque (OTT). To date, the experimental observation of OTT and its utilization for switching of antiferromagnetic layers in AATJs have remained elusive.

Here, we experimentally demonstrate OTT-induced switching of the CMO by flowing a vertical current through nanoscale $PtMn_3|MgO|PtMn_3$ AATJs. We find that current flowing through the junction drives the switching of the magnetic octupole moment without an external magnetic field, resulting in a resistance variation of the device with a TMR ratio up to 363% at room temperature. Using pulse width-dependent switching current density measurements, we estimate the energy barrier of the free layer to be approximately 39 kT.

We uncover a microscopic mechanism, confirmed by density functional theory (DFT) calculations, which underlies the observed OTT for a spin-neutral current traversing the antiferromagnetically alternating (111) Kagome planes. Specifically, OTT originates from the octupole-moment-driven imbalance in the contributions of different Mn sublattice atoms to the Bloch states extending across the tunnel barrier. This imbalance leads to a stronger intra-sublattice torque compared to inter-sublattice torque, thereby manifesting itself as a net staggered spin-transfer torque acting separately upon each sublattice. By incorporating such torques in micromagnetic simulations, we demonstrate OTT-driven switching consistent with experimental observations. The observation of OTT and its application in switching the CMO provide a pathway towards all-antiferromagnetic MRAM.

**Experimental observation of OTT-induced switching**

To observe the OTT switching of AATJs, we employed antiferromagnetic $PtMn_3$ as the active material of the tunnel junction. $PtMn_3$ is a cubic noncollinear antiferromagnet in which Mn spins reside in the Kagome (111)-plane with a 120° arrangement with respect to each other. While exchange and Dzyaloshinskii-Moriya interactions (DMI) stabilize the coplanar Kagome lattice, threefold uniaxial anisotropy in the (111)-plane allows two distinct ground state spin configurations: Mn spins pointing either all-inward or outward (Fig. 1a). When integrated into an AATJ structure, $PtMn_3$ has been reported to show a sizable TMR value at room temperature[11,13]. These characteristics render the $PtMn_3$-based AATJ a favorable platform for investigating OTT.

To this end, we deposited Pt (5 nm) / $PtMn_3$ (7 nm) / MgO (2 nm) / $PtMn_3$ (10 nm) / Pt (5 nm) / Ru (10 nm) stacks at room temperature by using a Canon ANELVA HC7100 physical vapor deposition (PVD) system. The films were sputter-deposited on a thermally oxidized 150 mm silicon wafer (Fig. 1a, Methods). The MgO thickness was chosen to exhibit resistance-area product (RA) values of approximately 1 to 10 Ωμm$^2$ to maximize the torque efficiency while maintaining moderate tunnelling resistance. Subsequently, the films were patterned into nanometer-sized circular pillars ranging from 50 to 200 nm in diameter, using electron (e-) beam lithography and dry etching for electrical measurements (Fig. 1b, Methods). Both the nanoscale lateral dimensions and low RA product are key ingredients for the realization of current-induced OTT switching.

We performed scanning transmission electron microscopy (STEM) on the 200 nm-diameter AATJ pillars to examine their crystallographic orientation (Fig. 1c). The results indicated the presence of a continuous MgO layer with a well-defined average thickness of approximately 2 nm despite some visible roughness, particularly at the bottom MgO-$PtMn_3$ interface. Both $PtMn_3$ layers were visibly polycrystalline, while the MgO layer did not indicate well-defined crystallization. We then analyzed the crystallinity of various grains through a fast Fourier transform (FFT) for both the top and bottom $PtMn_3$ layers, two of which are shown in the top and bottom panel of Fig. 1d, respectively. We found that both $PtMn_3$ layers had a predominant (111) orientation, as expected from films grown on a Pt seed layer, consistent with our previous study[13]. X-ray diffraction, X-ray fluorescence, and superconducting quantum interference device (SQUID) magnetometry measurements of the films are shown in Supplementary Note 1.

We next performed two-terminal current-induced switching measurements in our $PtMn_3$-based AATJs. Current pulses of varying amplitudes were applied to the AATJs, followed by a constant small reading current (Methods). The measurements were conducted at room temperature without any external magnetic field. Notably, we observed bidirectional switching with a sharp transition at the current density, $J$, of ~11 MA/cm$^2$ (Fig. 2a). When further increasing the current pulse amplitude, the tunneling resistance saturated, resulting in a TMR value of 363%. Different diameters of pillars from 50 to 200 nm also exhibited bidirectional switching at similar $J$ values (Extended Data Fig. 1). The observed switching behaviors were all reversible, repeatable, and showed clear saturation in both directions. We note that there were also AATJ devices that featured gradual multilevel switching (Extended Data Fig. 1), indicating a magnetic domain-driven switching mechanism.

To better understand this point, we separately performed nitrogen vacancy (NV) magnetometry measurements on different $PtMn_3$ films, which were sputtered on an MgO (100) substrate to obtain an epitaxial layer. The specific substrate orientation was used to make the stray field from the (001) surface of $PtMn_3$ detectable by our NV center tip, as it terminated with a ferromagnetic Mn atomic plane (Supplementary Note 2). The result reveals a magnetic domain size range of ~100-170 nm (Extended Data Fig. 2). It is thus reasonable to assume that only some of our devices, which have sizes comparable to or smaller than the expected domain size, would show abrupt and square-shaped switching. It is worth noting, however, that these NV microscopy results only provide an estimate of the domain size in our AATJs. The different crystal orientation which we had to use for the NV microscopy measurement, as well as inherent differences of epitaxial and polycrystalline samples, do not allow for a precise quantification of the transition size from multi- to single-domain behavior.

We next examined the statistical distribution of TMR within 50 devices including all sizes of pillars (Fig. 2b). Despite the variation in TMR values, current-induced switching was consistently present in our devices, with the highest observed TMR value reaching 363%. We note that crystallization of the MgO barrier was not required for the observation of large TMR in these devices, as further discussed in Supplementary Note 3. We also carried out current-induced switching measurements at different temperatures up to 400 K (Fig. 2c). We found that the current-induced switching and TMR both vanished at temperatures near 360 K, reappearing upon return to 300 K (Supplementary Note 4). This result is consistent with $PtMn_3$ undergoing a magnetic phase transition from a noncollinear to a collinear phase at elevated temperature[46,47] and thus further supports the fact that the switching of our AATJs arises from the noncollinear spin configuration of the $PtMn_3$ layers.

Lastly, we experimentally estimated the energy barrier (i.e., thermal stability factor) of an AATJ device by performing similar current-induced switching experiments with different pulse widths, ranging from 100 μs to 50 ms. The switching current density $J$ is plotted as a function of ln ($t_p/t_0$) in Fig. 2d, in which $t_p$ and $t_0$ refer to the current pulse width and the inverse of the attempt frequency, respectively. From a linear fit[48], we obtained an energy barrier of 39 kT at room temperature, consistent with our theoretical estimation (Methods and Extended Data Fig. 3). These bidirectional switching characteristics with moderately large thermal stability could facilitate applications such as all-antiferromagnetic magnetoresistive random-access memory.

It is worth noting that the Pt seed layer may provide a finite spin-orbit torque to the adjacent $PtMn_3$. However, the asymmetry of switching currents for opposite switching directions, which was consistently observed in all measured devices, rules out SOT as the primary switching mechanism. Supplementary Note 5 shows current-induced switching experiments using both in-plane currents (through the underlying Pt layer, i.e., in the SOT geometry) and out-of-plane currents (through the AATJ, i.e., in the OTT geometry) within the same structure. While OTT switching is observed consistent with the results of the main text, even significantly higher in-plane current densities in the Pt layer do not result in SOT-induced switching. This demonstrates that SOT is not the main mechanism responsible for switching in our devices.

**Microscopic mechanism of OTT**

To understand the origin of the current-induced switching in our AATJs, we analyze the electronic structure of $PtMn_3$ with its $T_{1g}$ CMO configuration[49]. In a crystalline AATJ operating in the ballistic transport regime, the transverse wavevector, $\boldsymbol{k}_{\|}$, is conserved due to the absence of diffusive scattering[29]. In this regime, we compute the $\boldsymbol{k}_{\|}$-resolved, operator-projected conduction channels, $N_{\|}^{\hat{O}}(\boldsymbol{k}_{\|})$, which count the number of Bloch states propagating along the transport direction (the $\langle 111 \rangle$ axis) weighted by their projection onto specific atomic sites, spin direction operators, or octupole operator, $\hat{O}$, at the Fermi energy (see Methods). In the following, we present a microscopic picture for the observed spin torque in terms of the sublattice-resolved spin polarizations and then discuss its equivalence to that based on the concept of the octupole polarization.

The atom-projected conduction channels, $N_{\|}^{(i)}(\boldsymbol{k}_{\|})$, exhibit sublattice-dependent anisotropic distribution across the Brillouin zone (Fig. 3a). Notably, each of the three high-symmetry Γ–K axes—related by 120° counterclockwise rotations—is dominated by one Mn sublattice, while the Γ–M axes are shared by all three sublattices. For a given sublattice, $i$, the Γ–M axis perpendicular to its dominant Γ–K axis shows a weaker contribution. When projecting onto site-spin states with quantization axis aligned to the local spin of $Mn_i$ (Fig. 3b), the site-spin-projected number of conduction channels exhibits a similar directional selectivity: Spin-majority ($N_{\|}^{(i\downarrow)}(\boldsymbol{k}_{\|})$) electrons dominate along the Γ–K axis of their corresponding sublattice and the two non-perpendicular Γ–M axes, whereas spin-minority ($N_{\|}^{(i\uparrow)}(\boldsymbol{k}_{\|})$) electrons dominate along the perpendicular Γ–M axis (Extended Data Fig. 4a,b). The resulting spin polarization, [$N_{\|}^{(i\sigma)}(\boldsymbol{k}_{\|}) = N_{\|}^{(i\downarrow)}(\boldsymbol{k}_{\|}) - N_{\|}^{(i\uparrow)}(\boldsymbol{k}_{\|})$], preserves the same anisotropic pattern (Fig. 3b). This decomposition of

the conduction channels and their spin polarization in the reciprocal space originates from the octupole configuration of the magnetic sublattices.

Schematics of the sublattice-resolved contribution of a select band (marked by blue and orange triangles) to the conduction channels, along with the corresponding projection onto atomic sites and spins, are presented in Fig. 3c. The direction (color) of the arrows represents the direction of the spin polarization (magnitude) of the contribution of the given sublattice, $Mn_i$, to the conduction channels at a given $\boldsymbol{k}_{||}$ along the Γ–K axes. The black (gray) color represents a large, $\beta_1$, (small, $\beta_2$) contribution from the given sublattice to the corresponding dominant (non-dominant) Γ–K axis. For a given $\boldsymbol{k}_{||}$ marked as $\boldsymbol{k}_1$ in Fig. 3c (dominant sublattice 1 contribution), considering the conservation of $\boldsymbol{k}_{||}$ during tunneling, the anisotropic distribution of $N_{||}^{(i)}(\boldsymbol{k}_{||})$ and $N_{||}^{(i\sigma)}(\boldsymbol{k}_{||})$ indicates that the electrons preferentially retain their sublattice identity and spin as they tunnel from one $PtMn_3$ layer to the other. The total spin polarizations of intra- and inter-sublattice currents are indicated by black and gray arrows in Fig. 3d, where $\boldsymbol{p}_i$ is the unit vector along the sublattice spin. The total spin polarization of the inter-sublattice current (pink arrow in Fig. 3d) is in the opposite direction of the intra-sublattice polarization, $\beta_2(\boldsymbol{p}_2 + \boldsymbol{p}_3) = -\beta_2\boldsymbol{p}_1$. Therefore, the total tunneling current to each sublattice, $Mn_i$, carries a net spin polarization of $(\beta_1 - \beta_2)\boldsymbol{p}_i$ (blue arrow in Fig. 3d) and induces an OTT of the form $\boldsymbol{T}_i \propto (\beta_1 - \beta_2)\boldsymbol{m}_i \times (\boldsymbol{m}_i \times \boldsymbol{p}_i)$. This torque resembles a sublattice-dependent STT and is a direct consequence of the octupolar order, as further discussed in the next section. Simulations of the dynamics using a three-sublattice Landau-Lifshitz-Gilbert (LLG) model, shown in Extended Data Fig. 5 and further discussed in Methods, demonstrate that this OTT mechanism can indeed give rise to deterministic and directional current-induced switching.

**Discussion and Outlook**

The proposed torque mechanism that is based on sublattice-resolved spin polarizations is fundamentally rooted in the underlying octupolar symmetry of the $PtMn_3$ magnetic structure, and thus can also be understood directly in terms of the octupole polarization. The $T_{1g}$ CMO is a symmetry-adapted description of the noncollinear spin structure of the three Mn sublattices in $PtMn_3$. Each $Mn_i$ sublattice contributes a spin moment oriented along one of the $\langle 2\bar{1}\bar{1} \rangle$ directions, forming a coplanar, non-collinear $T_{1g}$ octupole configuration in the (111) plane. In Extended Data Fig. 6a,b, we show the octupole-projected number of conduction channels—octupole minority, $N_{||}^{(T_+^\alpha)}(\boldsymbol{k}_{||})$, and octupole majority, $N_{||}^{(T_-^\alpha)}(\boldsymbol{k}_{||})$. It is seen that all conduction channels reveal a net octupole polarization, $N_{||}^{(T^\alpha)}(\boldsymbol{k}_{||}) = N_{||}^{(T_+^\alpha)}(\boldsymbol{k}_{||}) - N_{||}^{(T_-^\alpha)}(\boldsymbol{k}_{||})$ , along the octupole majority direction (Extended Data Fig. 6c). Importantly, the highest octupole polarization appears in the same space of the Brillouin zone as the sublattice-dominated states along the Γ–K axis which contribute strongly to each of the sublattice spin polarizations (Fig. 3a,b), underlining the direct connection of the sublattice-spin and cluster octupole pictures. Based on this analogy, we can conclude that the sublattice-resolved spin analysis provides a microscopic realization of the macroscopic cluster octupole moment. While the two descriptions of the observed torque are equivalent, the octupole-based description offers a symmetry-consistent formalism for describing the torque and transport behavior of AATJs and provides a direct analogy to the magnetization-based understanding of TMR and STT in conventional MTJs. We expect that our results will motivate the research community to develop a more complete theory of OTT and OTT-induced dynamics in AATJs.

Based on this understanding, the efficiency of the OTT could be further enhanced by engineering the band structure to maximize intra-sublattice hybridization of the states at the Fermi level, i.e., further enhancing the directional selectivity of the conduction channels in reciprocal space. Additionally, symmetry breaking or chemical substitution strategies that amplify the anisotropic sublattice character of the conduction channels may also serve to boost the torque efficiency. Other antiferromagnets within the $XMn_3$ family may provide a fruitful direction for further investigation. For example, $IrMn_3$ and $RhMn_3$ have the same crystal and magnetic noncollinear spin structure as $PtMn_3$ but with higher Néel temperatures, which may prove valuable in some applications.

At the device level, investigating the speed of OTT-induced dynamics is a fruitful direction for further investigation. For example, recent work has shown that in noncollinear antiferromagnets, large exchange fields can reduce the attempt time ($\tau_0$) by orders of magnitude compared to state-of-the-art perpendicular ferromagnets[50]. This suggests that thermally activated reversal via OTT in $PtMn_3$ may be significantly faster than in ferromagnetic MTJs.

It is also worth highlighting the role of surface termination of $PtMn_3$ on the observed switching characteristics. Even if both the top and bottom $PtMn_3$ layers contain a single lateral magnetic domain, surface roughness can generate regions with different relative surface termination on the two sides of the tunnel barrier. As discussed in detail in Methods and Extended Data Fig. 7, these variations in surface termination between the two electrodes lead to local variations of the TMR and OTT efficiency, which can contribute to the gradual switching observed in some of the measured devices. A micromagnetic simulation of the size dependence of domain wall nucleation and dynamics in the presence of OTT is presented in Supplementary Note 6.

Overall, the observed current-induced OTT switching of nanoscale $PtMn_3$|MgO|$PtMn_3$ AATJs and the associated large room-temperature TMR represent an important milestone in antiferromagnetic spintronics. Our results clearly demonstrate that AATJs can be efficient both in write-in and read-out operations and thus open a new route towards a variety of antiferromagnetic

spintronic devices. Nanoscale AATJs can, for example, be used to realize magnetic random-access memory arrays with higher resilience to external magnetic fields while fully eliminating dipole interaction between neighboring bits. Beyond memory, AATJs could be engineered to exhibit smaller energy barriers that would make them useful in applications such as entropy sources and probabilistic bits. Finally, AATJs open opportunities for fully-electrical access to the ultrafast exchange-dominated dynamics of noncollinear antiferromagnets, e.g., through experiments to realize OTT-based high-frequency rectification and OTT-induced oscillations. Given the relative accessibility of this material system and its growth in industry-standard deposition equipment, we expect our results to rapidly drive progress in both fundamental and applied spintronics experiments.

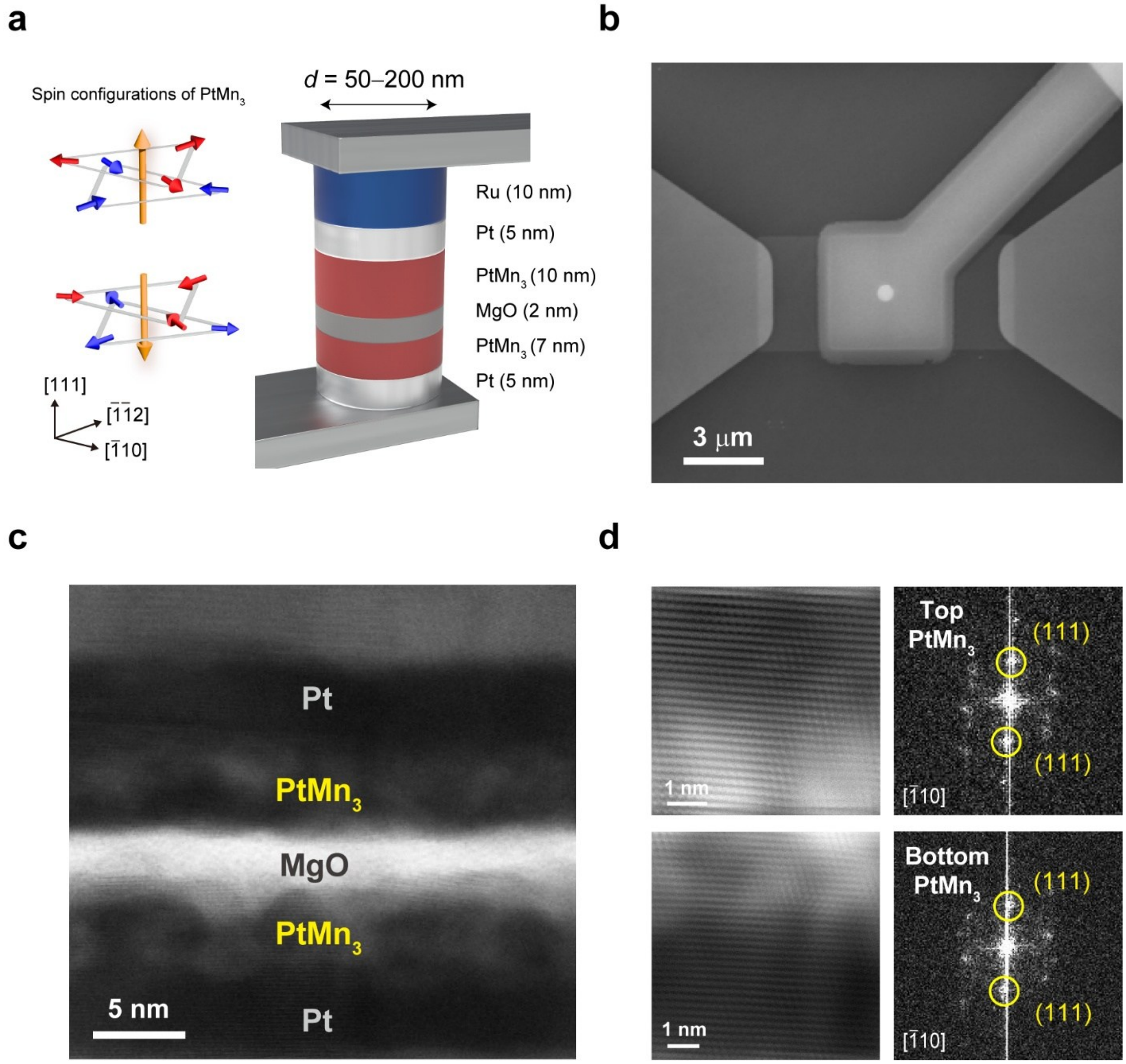

**Fig. 1| Device structure of all-antiferromagnetic tunnel junctions. a**, Spin configurations (red and blue) and associated cluster magnetic octupole (orange) in $PtMn_3$ (left). Material stack of AATJs used in the experiment (right). **b**, Scanning electron microscopy image of an AATJ device. **c,** Cross-sectional scanning transmission electron microscopy (STEM) image of Pt (5 nm) / $PtMn_3$ (7 nm) / MgO / $PtMn_3$ (10 nm) / Pt (5 nm) / Ru (10 nm) AATJ. **d**, STEM images (left) and the corresponding fast Fourier transformed (FFT) patterns (right) of the top $PtMn_3$ (top) and bottom $PtMn_3$ (bottom), respectively, in which the predominant (111) orientation is observed.

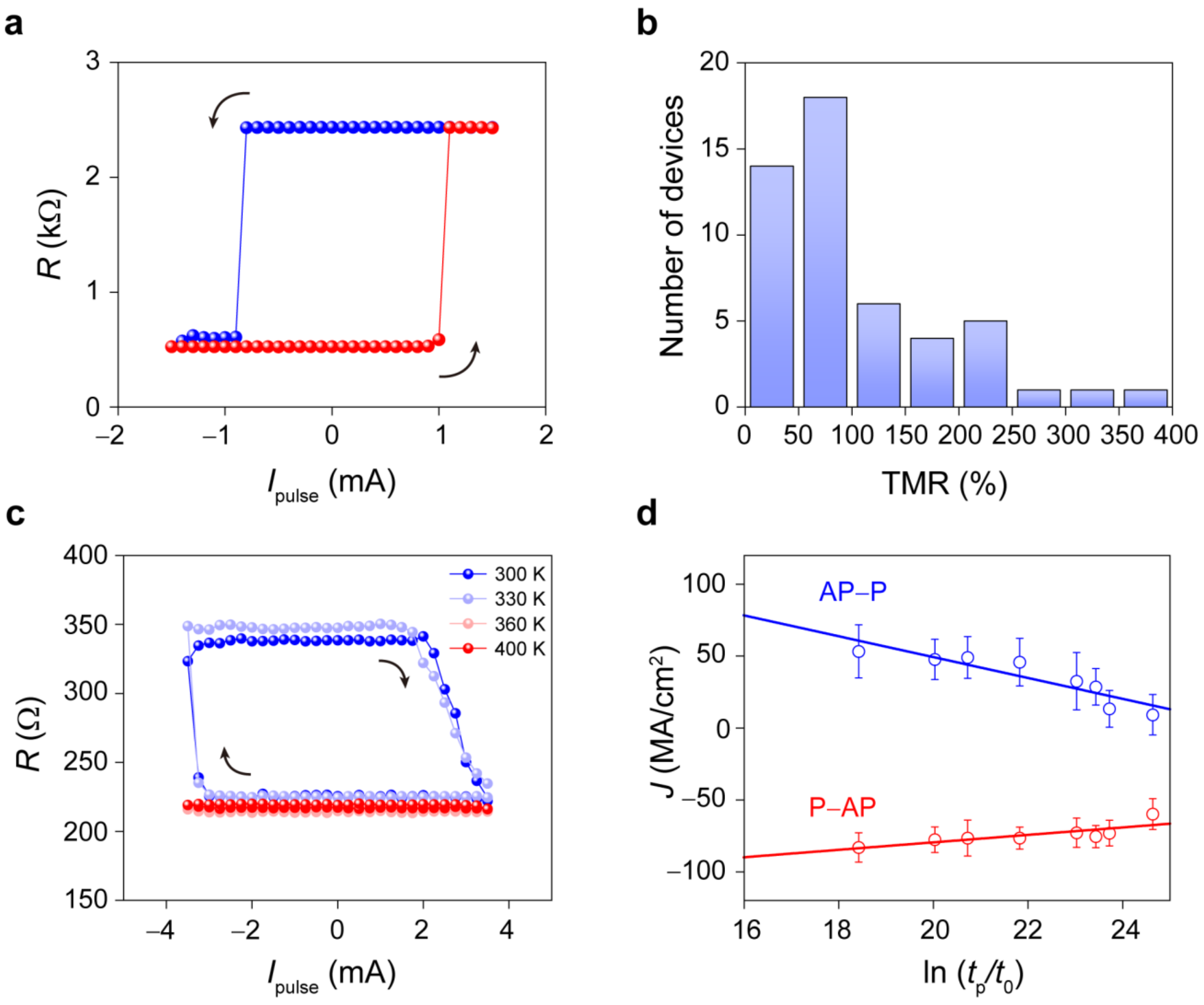

**Fig. 2| Two-terminal current-induced switching of AATJs. a**, Current-induced switching measurements on a 100 nm-diameter AATJ device. **b**, Statistical distribution of TMR from current-induced switching. **c,** Temperature-dependent switching curves measured from 300 K to 400 K. **d**, Switching current density (*J*) plotted as a function of ln ($t_p/t_0$), obtained from pulse width-dependent switching measurements on a 100 nm-diameter AATJ device. We assumed $t_0$ = 1 ps.

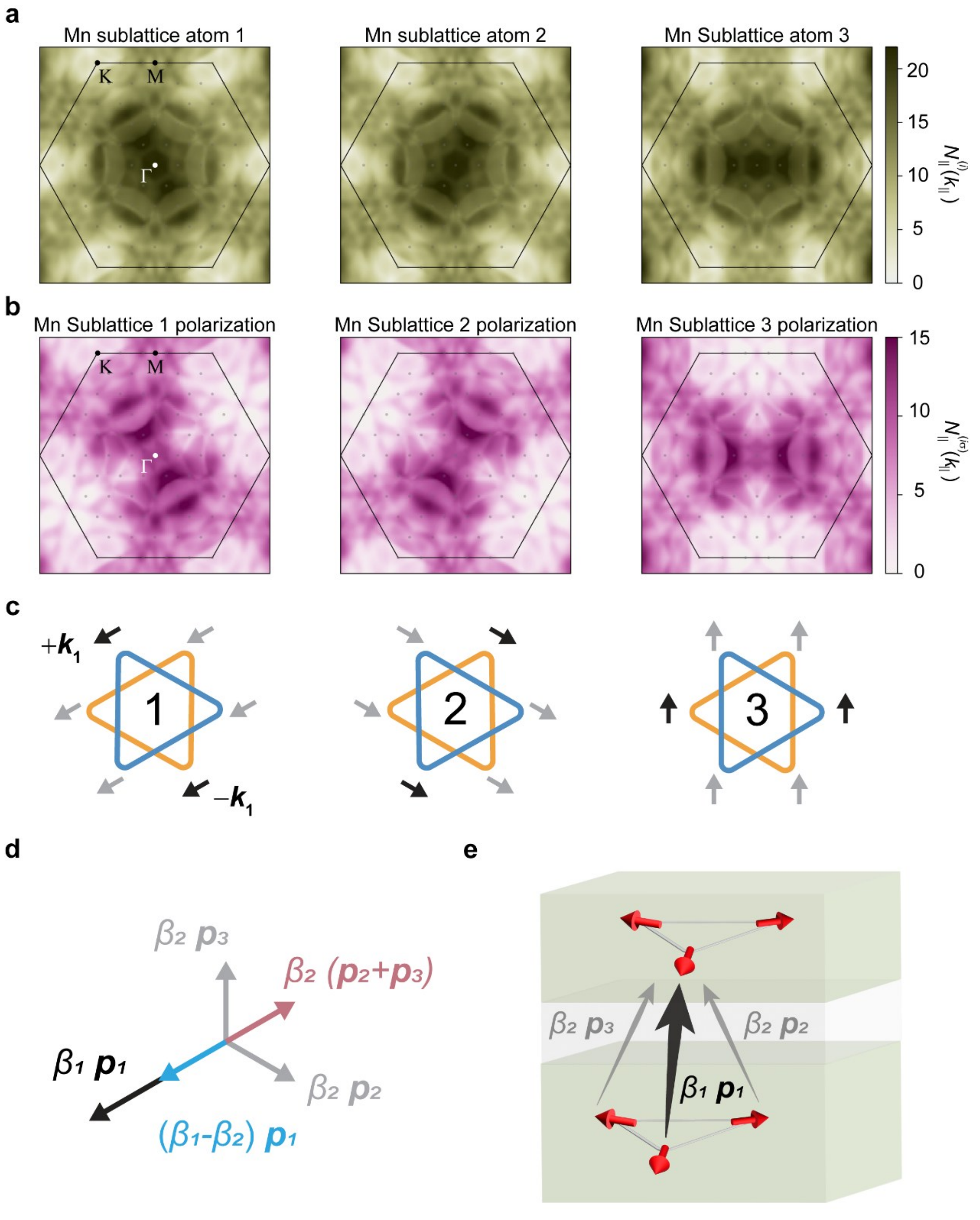

**Fig. 3| Sublattice-projected conduction channels as a function of k$_{||}$ for PtMn$_3$. a**, **b**, Number of conduction channels, projected onto the sublattice Mn atomic site **(a)** and the respective spin direction of the corresponding sublattice **(b)**. **c**, Schematic representation of the Fermi surface and

associated atomic and spin projection along the Γ-K directions for different sublattices. Black (gray) arrows represent high (low) number of conduction channels. **d**, $\beta_1$ ($\beta_2$), with the corresponding spin polarization along the sublattice spin direction along the Γ-K directions. Black (gray) arrows represent the polarization of the intra (inter)-sublattice tunneling current to sublattice $Mn_1$ from the indicated $\boldsymbol{k}_1$ points. **e**, Black (gray) arrows represent intra (inter)-sublattice tunneling current.

## References

1. Moodera, J. S., Kinder, L. R., Wong, T. M. & Meservey, R. Large magnetoresistance at room temperature in ferromagnetic thin film tunnel junctions. *Phys. Rev. Lett.* **74**, 3273-3276 (1995).
2. Miyazaki, T. & Tezuka, N. Giant magnetic tunneling effect in Fe/$Al_2O_3$/Fe junction. *J. Magn. Magn. Mater.* **139**, L231-L234 (1995).
3. Tsymbal, E. Y., Mryasov, O. N. & LeClair, P. R. Spin-dependent tunnelling in magnetic tunnel junctions. *J. Phys.: Condens. Matter* **15**, R109-R142 (2003).
4. Parkin, S. S. et al. Giant tunnelling magnetoresistance at room temperature with MgO (100) tunnel barriers. *Nat. Mater.* **3**, 862-867 (2004).
5. Yuasa, S., Nagahama, T., Fukushima, A., Suzuki, Y. & Ando, K. Giant room-temperature magnetoresistance in single-crystal Fe/MgO/Fe magnetic tunnel junctions. *Nat. Mater.* **3**, 868-871 (2004).
6. Slonczewski, J. C. Current-driven excitation of magnetic multilayers. *J. Magn. Magn. Mater.* **159**, L1-L7 (1996).
7. Berger, L. Emission of spin waves by a magnetic multilayer traversed by a current. *Phys. Rev. B* **54**, 9353 (1996).
8. Katine, J. A., Albert, F. J. & Buhrman, R. A. Current-driven magnetization reversal and spin-wave excitations in Co/Cu/Co pillars. *Phys. Rev. Lett.* **84**, 3149 (2000).
9. Diao, Z. et al. Spin transfer switching and spin polarization in magnetic tunnel junctions with MgO and $AlO_x$ barriers *Appl. Phys. Lett.* **87**, 232502 (2005).
10. Dong, J. et al. Tunneling magnetoresistance in noncollinear antiferromagnetic tunnel junctions. *Phys. Rev. Lett.* **128**, 197201 (2022).
11. Qin, P. et al. Room-temperature magnetoresistance in an all-antiferromagnetic tunnel junction. *Nature* **613**, 485-489 (2023).
12. Chen, X. et al. Octupole-driven magnetoresistance in an antiferromagnetic tunnel junction. *Nature* **613**, 490-495 (2023).
13. Shi, J. et al. Electrically controlled all-antiferromagnetic tunnel junctions on silicon with large room-temperature magnetoresistance. *Adv. Mater.* **36**, e2312008 (2024).
14. Wolf, S. A. et al. Spintronics: a spin-based electronics vision for the future. *Science* **294**, 1488-1495 (2001).

15. Tulapurkar, A. A. et al. Spin-torque diode effect in magnetic tunnel junctions. *Nature* **438**, 339-342 (2005).
16. Železný, J. et al. Relativistic Néel-order fields induced by electrical current in antiferromagnets. *Phys. Rev. Lett.* **113**, 157201 (2014)
17. Wadley, P. et al. Electrical switching of an antiferromagnet. *Science* **351**, 587-590 (2016)
18. Baltz, V. et al. Antiferromagnetic spintronics. *Rev. Mod. Phys.* **90** (2018).
19. Olejnik, K. et al. Terahertz electrical writing speed in an antiferromagnetic memory. *Sci. Adv.* **4**, eaar3566 (2018).
20. Šmejkal, L., Mokrousov, Y., Yan, B. & MacDonald, A. H. Topological antiferromagnetic spintronics. *Nat. Phys.* **14**, 242-251 (2018).
21. DuttaGupta, S. et al. Spin-orbit torque switching of an antiferromagnetic metallic heterostructure. *Nat. Commun.* **11**, 5715 (2020).
22. Shi, J. et al. Electrical manipulation of the magnetic order in antiferromagnetic PtMn pillars. *Nat. Electron.* **3**, 92-98 (2020).
23. Tsai, H. et al. Electrical manipulation of a topological antiferromagnetic state. *Nature* **580**, 608-613 (2020).
24. Arpaci, S. et al. Observation of current-induced switching in non-collinear antiferromagnetic $IrMn_3$ by differential voltage measurements. *Nat. Commun.* **12**, 3828 (2021).
25. Du, A. et al. Electrical manipulation and detection of antiferromagnetism in magnetic tunnel junctions. *Nat. Electron.* **6**, 425–433 (2023).
26. Shao, D.-F. & Tsymbal, E. Y. Antiferromagnetic tunnel junctions for spintronics. *npj Spintronics* **2** (2024).
27. Ahn, K.-H., Hariki, A., Lee, K.-W. & Kuneš, J. Antiferromagnetism in $RuO_2$ as *d*-wave Pomeranchuk instability. *Phys. Rev. B* **99**, 184432 (2019).
28. Yuan, L.-D., Wang, Z., Luo, J.-W., Rashba, E. I. & Zunger. A. Giant momentum-dependent spin splitting in centrosymmetric low-Z antiferromagnets. *Phys. Rev. B* **102**, 014422 (2020).
29. Shao, D.-F., Zhang, S.-H., Li. M., Eom, C.-B. & Tsymbal, E. Y. Spin-neutral currents for spintronics. *Nat. Commun.* **12**, 7061 (2021).
30. Šmejkal, L., Sinova, J. & Jungwirth, T. Emerging research landscape of altermagnetism. *Phys. Rev. X* **12** (2022).

31. Feng, Z. et al. An anomalous Hall effect in altermagnetic ruthenium dioxide. *Nat. Electron.* **5**, 735-743 (2022).
32. Krempaský, J. et al. Altermagnetic lifting of Kramers spin degeneracy. *Nature* **626**, 517-522 (2024).
33. Chen, H., Niu, Q. & MacDonald, A. H. Anomalous Hall effect arising from noncollinear antiferromagnetism. *Phys. Rev. Lett.* **112**, 017205 (2014).
34. Higo, T. et al. Perpendicular full switching of chiral antiferromagnetic order by current. *Nature* **607**, 474-479 (2022).
35. Rimmler, B. H., Pal, B. & Parkin, S. S. P. Non-collinear antiferromagnetic spintronics. *Nat. Rev. Mater.* **10**, 109-127 (2024).
36. Takeuchi., Y. et al. Electrical coherent driving of chiral antiferromagnet. *Science* **389**, 830-834 (2025).
37. Nakatsuji, S., Kiyohara, N. & Higo, T. Large anomalous Hall effect in a non-collinear antiferromagnet at room temperature. *Nature* **527**, 212-215 (2015).
38. Ikhlas, M. et al. Large anomalous Nernst effect at room temperature in a chiral antiferromagnet. *Nat. Phys.* **13**, 1085-1090 (2017).
39. Higo, T. et al. Large magneto-optical Kerr effect and imaging of magnetic octupole domains in an antiferromagnetic metal. *Nat. Photonics* **12**, 73-78 (2018).
40. Chou, C. T. et al. Large spin polarization from symmetry-breaking antiferromagnets in antiferromagnetic tunnel junctions. *Nat. Commun.* **15**, 7840 (2024).
41. Gurung, G., Elekhtiar, M., Luo, Q. Q., Shao, D.-F. & Tsymbal, E. Y. Nearly perfect spin polarization of noncollinear antiferromagnets. *Nat. Commun.* **15**, 10242 (2024).
42. Tserkovnyak, Y. & Mecklenburg, M. Electron transport driven by nonequilibrium magnetic textures. *Phys. Rev. B* **77** (2008).
43. Brataas, A., Tserkovnyak, Y., Bauer, G. E. W. & Kelly, P. J. Spin pumping and spin transfer. In *Spin Current* 87–135 (eds Maekawa, S., Valenzuela, S. O., Saitoh, E. & Kimura, T.) (Oxford Univ. Press, Oxford, 2012).
44. Ghosh, S., Manchon, A. & Zelezny, J. Unconventional Robust Spin-Transfer Torque in Noncollinear Antiferromagnetic Junctions. *Phys. Rev. Lett.* **128**, 097702 (2022).

45. Liu, S. et al. $Mn_3SnN$-based antiferromagnetic tunnel junction with giant tunneling magnetoresistance and multi-states: design and theoretical validation. *Adv. Sci.* **12**, e02985 (2025).
46. Ikeda, T. & Tsunoda, Y. Spin fluctuations in an octahedral antiferromagnet $Mn_3Pt$ alloy. *J. Phys. Soc. Jpn.* **72**, 2614-2621 (2003).
47. Liu, Z. Q. et al. Electrical switching of the topological anomalous Hall effect in a non-collinear antiferromagnet above room temperature. *Nat. Electron.* **1**, 172-177 (2018).
48. Koch, R. H., Katine, J. A. & Sun, J. Z. Time-resolved reversal of spin-transfer switching in a nanomagnet. *Phys. Rev. Lett.* **92**, 088302 (2004).
49. Suzuki, M. T., Koretsune, T., Ochi, M. & Arita, R. Cluster multipole theory for anomalous Hall effect in antiferromagnets. *Phys. Rev. B* **95**, 094406 (2017).
50. Konakanchi, S. T., et al. Electrically tunable picosecond-scale octupole fluctuations in chiral antiferromagnets. *Phys. Rev. Lett.* **135**, 136704 (2025).

# Methods

## Sample growth and device fabrication

To fabricate the $PtMn_3|MgO|PtMn_3$-based AATJs for switching experiments, Pt (5 nm) / $PtMn_3$ (7 nm) / MgO / $PtMn_3$ (10 nm) / Pt (5 nm) / Ru (10 nm) film stacks were grown by sputter deposition on thermally oxidized 150 mm silicon wafers using a Canon ANELVA HC7100 physical vapor deposition (PVD) system. All metallic layers were deposited by dc magnetron sputtering. The $PtMn_3$ layer was deposited by co-sputtering of Pt and Mn targets. The MgO layer was deposited using RF sputtering. All the layers were deposited at room temperature without post-deposition annealing. After deposition, the samples were patterned into MTJ pillars with a diameter ranging from 50 nm to 200 nm using electron beam lithography and dry etching.

## Electrical measurements

Two-terminal electrical measurements were conducted using a Keithley 6221 current source and a Keithley 2182a nanovoltmeter. Pulse delta sweep mode was employed to measure the switching curves. The pulse width of the current pulse and constant reading current were both 1 ms unless otherwise specified. The temperature-dependent switching curves were measured by using a Lakeshore CRX-VF probe station.

## $PtMn_3$ for NV magnetometry measurements

A $PtMn_3$ (10 nm) / Pt (5 nm) sample was grown by DC magnetron sputtering on a single-crystalline MgO (100) layer in a PVD system with a base pressure of $5\times10^{-9}$ Torr. An RF plasma cleaning step was performed, and the substrate was then annealed at 850˚C for 2 hours under $O_2$ pressure of 20 mT to reconstruct an atomically flat surface for epitaxial growth. The $PtMn_3$ layer was deposited by co-sputtering of Pt and Mn targets.

**NV magnetometry measurements**

The NV center has a ground state spin triplet, $m_s = 0, \pm 1$. A small external magnetic field lifts the degeneracy of $m_s = \pm 1$. The spin-state can be optically initialized into $m_s = 0$ via off-resonance green laser excitation and read out via spin-dependent photoluminescence as $m_s = 0$ emits more photons per second than $m_s = \pm 1$. The NV ground state Hamiltonian has a Zeeman term that governs the energy of the $m_s = 0 \leftrightarrow m_s = -1$ and $m_s = 0 \leftrightarrow m_s = +1$ spin transitions dependence on external magnetic fields. Here, optically detected magnetic resonance (ODMR) measurements are performed using an applied microwave magnetic field that switches between two frequencies on opposite sides of the $m_s = 0 \leftrightarrow m_s = -1$ transition. The difference of the fluorescence level at each pixel is converted into the external magnetic field from the sample (dual-iso-B mode). Scanning NV center magnetometry uses a diamond probe containing a single NV center near the surface that is scanned across the sample. The ODMR measurements are done as the probe is scanned laterally at a fixed, small height $d = 50$ nm (sample-to-probe distance) with a step size of 20 nm, resulting in a 2D magnetic field image at a single plane of the stray fields from a magnetic sample. To remove the high-frequency noise in the measurement due to scanning lines, we apply a Gaussian filter with standard deviation of $\sigma = 16$ nm. As $\sigma < d$, this filter acts like an identity operator with minimal distortion of the raw data. This is equivalent to a filter in the Fourier space with a standard deviation of $\sigma_k = \frac{2}{5} k_{\mathrm{Nyq}}$= 62.5 rad/µm, where $k_{\mathrm{Nyq}}$ is the Nyquist wavenumber corresponding to the scanning resolution.

**Scanning transmission electron microscopy (STEM)**

Cross-sectional lamella samples processed from a 100 nm AATJ device were prepared by focused ion beam (FIB) using the Thermo Scientific Helios 5 Hydra CX DualBeam Plasma FIB/SEM.

Bulk-out, lift-out, and thinning processes were performed using a 30 kV Xe+ ion beam. Final cleaning steps were performed at 8 kV and 5 kV to remove amorphous materials on the surface. Samples were then plasma cleaned using a South Bay Technology PC-2000 Plasma Cleaner with Ar plasma at 20 ~ 40 W RF power for 15 s at 150 mTorr to further remove residual contamination. The final sample thicknesses were approximately 50-80 nm. Scanning transmission electron microscopy (STEM) data were collected on an aberration-corrected JEOL JEM-ARM200CF STEM operating at 200 kV. The STEM convergence angle, HAADF collection angle, and ABF collection angle were 27 mrad, 90-370 mrad, and 23 mrad, respectively. Data processing (fast Fourier transform and denoising) was performed using the Gatan Microscopy Suite (GMS) software.

**Calculation of the energy barrier in a $PtMn_3$ free layer**

$PtMn_3$ has an $L1_2$ crystal structure with Mn atoms at the face centers and Pt atoms at the corners of the cubic cell. The cubic anisotropy is locally broken for each of the Mn spins, which when combined with the antiferromagnetic exchange, results in the three Mn spins forming a triangle as shown in Fig. 1a within the (111) Kagome plane[51]. In the micromagnetic continuum limit, the state of a $PtMn_3$ nanomagnet may be represented by three classical magnetization vectors, $\boldsymbol{M}_i = M_s\boldsymbol{m}_i$. Each vector corresponds to a pair of ferromagnetically locked sublattice spins, with $M_s$ being the saturation magnetization[50,52]. A minimal model for the free energy density of such a nanomagnet in the monodomain limit may be expressed as[52-54],

$$F = J(\boldsymbol{m}_1 \cdot \boldsymbol{m}_2 + \boldsymbol{m}_2 \cdot \boldsymbol{m}_3 + \boldsymbol{m}_3 \cdot \boldsymbol{m}_1) + D[\boldsymbol{z} \cdot (\boldsymbol{m}_1 \times \boldsymbol{m}_2 + \boldsymbol{m}_2 \times \boldsymbol{m}_3 + \boldsymbol{m}_3 \times \boldsymbol{m}_1)]$$

$$-K_{eff}[(\boldsymbol{e}_1 \cdot \boldsymbol{m}_1)^2 + (\boldsymbol{e}_2 \cdot \boldsymbol{m}_2)^2 + (\boldsymbol{e}_3 \cdot \boldsymbol{m}_3)^2]. \quad (1)$$

Here, $J > 0$, $D < 0$ and $K_{\mathrm{eff}} > 0$ parameterize the antiferromagnetic exchange, Dzyaloshinskii-Moriya interaction (DMI), and effective uniaxial anisotropy energies, respectively, and $\boldsymbol{e}_i$ is the effective easy axis for the sublattice $\boldsymbol{m}_i$ in the Kagome plane[51]. Here, we neglect any small deviations of the Mn spins out of the Kagome plane in equilibrium. Extended Data Fig. 3a shows the two equilibrium configurations of the unit cell of $PtMn_3$. Electrical signatures of the anomalous Hall effect and tunneling magnetoresistance in $PtMn_3$ have been shown to correspond to ferroic order of the so-called octupole moment[49] which tracks the rotations of the $PtMn_3$ spin motif in the Kagome plane. The hierarchy of energy scales of the system, $J \gg |D| \gg K$, allows for a perturbative expansion of the free energy. Following the literature[50,55,56], we perturbatively expand Eq. (1) in normal mode coordinates of $PtMn_3$ in the exchange limit, i.e., when $D = K = 0$. These normal mode coordinates correspond to spin cantings along $x$, $y$ or $z$ directions and the corresponding rigid rotations of the $PtMn_3$ spin motif in the $y$-$z$, $x$-$z$ or $x$-$y$ planes. We identify that the rotation of the octupole moment within the Kagome plane ($x$-$y$ plane) corresponds to the lowest energy mode and adiabatically eliminates the other two high energy modes to obtain an effective free energy density for the in-plane mode as,

$$F_{xy} = \frac{9J+3\sqrt{3}D}{2} m_z^2 + 3K_{\mathrm{eff}}\cos^2\phi. \quad (2)$$

Here, $m_z = \cos[(\theta_1 + \theta_2 + \theta_3)/3]$ is the $z$ component and $\phi = (\phi_1 + \phi_2 + \phi_3 - 2\pi)/3$ is the azimuthal angle of the $PtMn_3$ spin motif. For a single-domain nanomagnet of volume $V$, the energy barrier for octupole rotations within the Kagome plane can then be read as $\Delta = 3K_{\mathrm{eff}}V$. Furthermore, using an in-house SPICE-based Landau-Lifshitz-Gilbert (LLG) equation simulator[57], we numerically calculate the energy barrier for the free energy given in Eq. (1) by simulating the $PtMn_3$ unit cell with three coupled sublattices. Extended Data Fig. 3b shows the energy of a 100 nm × 7 nm nanomagnet as a function of the azimuthal angle in the Kagome plane. The energies are obtained by rigidly rotating the $PtMn_3$ spin motif in Extended Data Fig. 3a by the angle $\phi$, locally relaxing the spins and numerically calculating the energy using Eq. (1). For the simulations, we use the parameters[13,53] $J = 280 \times 10^6$ J/m$^3$, $D = 0.1J$, and $M_s = 1110$ emu/cm$^3$. Extended Data Fig. 3b shows an excellent agreement between the numerically evaluated energy barrier and that predicted by Eq. (2). Matching the numerics to $\Delta = 39$ kT evaluated from OTT switching currents, the effective anisotropy for our samples may be extracted as $K_{eff} = 1$ kJ/m$^3$, which is consistent with the values reported in the literature[11,58,59].

**First-principles calculations**

All conduction channel calculations were performed using density functional theory (DFT) as implemented in the Vienna Ab initio Simulation Package (VASP)[60]. The projector augmented wave (PAW) method[61] was employed to describe the core-valence electron interactions for the cubic $PtMn_3$ primitive cell. For the electronic structure calculations, the generalized gradient approximation (GGA) with the Perdew-Burke-Ernzerhof (PBE) exchange-correlation functional[62] was utilized. To accurately describe the strongly correlated Mn d-electrons, a Hubbard-U correction of U = 2 eV was applied within the PBE+U framework[63]. The plane wave kinetic energy cutoff was set to 500 eV. The electronic convergence criteria for self-consistency were set to $10^{-6}$ eV. Geometric optimizations were performed until the Hellmann-Feynman forces on each atom were less than 0.01 eV/Å. The Brillouin zone was sampled by a Γ-centered 11×11×11 k-mesh. Gaussian smearing was used with $\sigma = 0.01$ eV. Relativistic spin-orbit coupling (SOC) was included in the calculation.

The operator-projected number of conduction channels is defined as

$$N_{||}^{(\hat{O})}(\boldsymbol{k}_{||}) = \sum_n \int dk_\perp\, \delta(\varepsilon_{n\boldsymbol{k}} - \varepsilon_f)\, |v_{n\boldsymbol{k}}^\perp| \langle n\boldsymbol{k}|\hat{O}|n\boldsymbol{k}\rangle \quad (3)$$

where $\boldsymbol{k}_{||}$ are the $k$-components in the (111)-plane, $\boldsymbol{k}_\perp$ is the $k$-component perpendicular to the (111)-plane (along the ⟨111⟩ direction), $\varepsilon_{n\boldsymbol{k}}$, is the energy of the $n^{\text{th}}$ band at wavevector $\boldsymbol{k}$, $v_{n\boldsymbol{k}}^\perp = \frac{\partial \varepsilon_{n\boldsymbol{k}}}{\hbar \partial k_\perp}$ is the group velocity along the ⟨111⟩ direction and $\langle n\boldsymbol{k}|\hat{O}|n\boldsymbol{k}\rangle$ is the expectation value of a general spin-atom operator $\hat{O}$. To calculate the conduction channel, we constructed a Wannier model[64] for the primitive cubic $PtMn_3$ unit cell, considering 48 orbitals, including the s- and d-orbitals of the Pt atom, and all s- and d-orbitals of the three Mn atoms. The primitive cell model was then unfolded to a hexagonal (111) surface supercell. The value of $N_{||}^{(\hat{O})}(\boldsymbol{k}_{||})$ was then evaluated on a 120×120×80 $k$-mesh interpolated from the surface Wannier model and integrated over $\boldsymbol{k}_\perp$. The spin polarization for the Bloch states was extrapolated from VASP calculations and interpolated onto the 120×120×80 $k$-mesh. The projected spin polarizations onto the three sublattices ($i$=1–3) were calculated using the following equations for spin quantization axes, $\hat{\sigma}_1 = 2\hat{\sigma}_x - \hat{\sigma}_y - \hat{\sigma}_z$, $\hat{\sigma}_2 = -\hat{\sigma}_x + 2\hat{\sigma}_y - \hat{\sigma}_z$ and $\hat{\sigma}_3 = -\hat{\sigma}_x - \hat{\sigma}_y + 2\hat{\sigma}_z$, respectively. The site projection was calculated directly by evaluating the expectation value of atomic site operator, $\hat{P}_i = |\text{Mn}_i\rangle\langle \text{Mn}_i|$ using the Wannier model. The octupole projected conduction channels were calculated by evaluating the expectation value of the cluster magnetic octupole operator defined in the literature for the $T_{1g}$ CMO configuration[49] [in the (111)-plane, $\hat{T}_{z'}^{\alpha} = \frac{1}{\sqrt{3}}(\hat{T}_x^{\alpha} + \hat{T}_y^{\alpha} + \hat{T}_z^{\alpha})$, where $\hat{T}_x^{\alpha}$, $\hat{T}_y^{\alpha}$, and $\hat{T}_z^{\alpha}$ are the components of the octupole operator defined in the conventional cell], and the total Mn atom projector, $\hat{P}_{\text{Mn}} = \sum_i |\text{Mn}_i\rangle\langle \text{Mn}_i|$. The cluster octupole minority and majority projected number of conduction channels were calculated as, $N_{||}^{(T_+^{\alpha})} = \frac{1}{2}\{N_{||}^{(\hat{P}_{\text{Mn}})} + N_{||}^{(\hat{T}_{z'}^{\alpha})}\}$ and $N_{||}^{(T_-^{\alpha})} = \frac{1}{2}\{N_{||}^{(\hat{P}_{\text{Mn}})} - N_{||}^{(\hat{T}_{z'}^{\alpha})}\}$, respectively.

**Effect of interface termination on the TMR distribution of AATJ devices**

The experimentally observed spread in TMR values, shown in Fig. 2b, is understandable given the microstructural characteristics of our stacks: while the magnetic electrodes are highly (111)-textured, the tunnel barrier is amorphous. As a result, individual junctions experience variations in the local atomic configuration at the crystalline–amorphous interface, including differences in interface termination, short-range order within the barrier, and local fluctuations in the effective barrier height or thickness. These factors are expected to affect spin-dependent transmission, and their device-to-device variation naturally leads to a statistical distribution of TMR values.

To illustrate one possible origin of the TMR distribution, we modeled the effect of interface termination on the TMR response. Specifically, we performed theoretical calculations of the electron transmission across AATJs in which two antiferromagnetic $PtMn_3$ layers are separated by a vacuum gap, and different interface terminations are simulated by introducing a lateral shift $\boldsymbol{d}_{\parallel}$ of the top $PtMn_3$ layer relative to the bottom (Extended Data Fig. 7a). This approach allowed us to mimic various stacking configurations, including those where the magnetic structures at the two interfaces do not align and cases where the magnetic layer at one interface effectively encounters a layer of different registry on the opposite side. A vacuum barrier was used to simplify the modeling and isolate the effect of interface termination.

Extended Data Figs. 7b and 7c show the calculated $\boldsymbol{k}_{\parallel}$- resolved transmission for parallel $T_P(\boldsymbol{k}_{\parallel})$ (b) and antiparallel $T_{AP}(\boldsymbol{k}_{\parallel})$ (c) alignment of antiferromagnetic $PtMn_3$ layers when $\boldsymbol{d}_{\parallel} = 0$, i.e., the magnetic moments of the top interface are perfectly aligned with the magnetic moments of the bottom interface. Such magnetic moment matching at the interfaces, combined with that in the bulk electrodes, makes $T_P(\boldsymbol{k}_{\parallel})$ much larger than $T_{AP}(\boldsymbol{k}_{\parallel})$ for almost all $\boldsymbol{k}_{\parallel}$, resulting in total $TMR = (T_P - T_{AP})/T_{AP}$ of about 600%. Shifting the top electrode with respect to the bottom, as described by the parameter $\boldsymbol{d}_{\parallel}$, reduces $T_P$, while $T_{AP}$ does not change much. As a result, the TMR decreases but remains positive for most lateral displacements $\boldsymbol{d}_{\parallel}$ except near shifts of approximately half a unit cell, where the minimum TMR is about –100% (Extended Data Fig. 7d).

An important observation that follows from this analysis is that the TMR is primarily determined by the bulk magnetic structure of the electrodes. Magnetic moment matching in the P alignment, contrasted to the mismatch in the AP alignment, largely determines the sign of the TMR, although the interface termination still plays a role. We would like to point out that the interface termination is constrained by the underlying bulk structure of $PtMn_3$ and cannot adopt arbitrary magnetic configurations. Due to bulk $PtMn_3$ exhibiting globally broken time-reversal symmetry, an exact match of both the magnetic and atomic structures at the two interfaces is not possible for the AP-aligned electrodes. This is why the distribution of TMR in Extended Data Fig. 7d is strongly biased toward positive values.

To enable comparison to our experimentally measured values, we consider the absolute value of the TMR ratio, defined as $TMR_a = |T_P - T_{AP}|/\min(T_P, T_{AP})$. In Extended Data Fig. 7e, we plot the distribution of the absolute TMR defined in this manner as obtained from our calculations, where the probability is represented by the normalized density of TMR values. Remarkably, this simple analysis captures the qualitative features of the experimental distribution shown in Fig. 2b of the main text. We conclude, therefore, that the observed broad distribution of TMR values is not unexpected for junctions lacking full crystallinity in the stack. Improving the TMR uniformity would require exploring growth conditions that promote higher crystallinity and better control over surface termination.

## Micromagnetic simulations

We assume that the intra- and inter-sublattice torques have the same form of Slonczewski STT in MTJs. The torque from the $j$-th sublattice in the fixed layer exerted on the $i$-th sublattice in the free layer is given by

$$\boldsymbol{T}_{ij} = \beta_{ij}\boldsymbol{\tau}_{ij} = \beta_{ij}\mathrm{A_{ij}}(J)\boldsymbol{m}_i \times (\boldsymbol{m}_i \times \boldsymbol{p}_j), \tag{4}$$

where $\mathrm{A_{ij}}(J) = \frac{g_e\mu_B}{eM_sd}2\eta J$ is the magnitude of the torque, $\beta_{ij}$ is the torque efficiency ($\beta_{ij} = \beta_1$ for intra-sublattice and $\beta_{ij} = \beta_2$ for inter-sublattice torque), $\boldsymbol{p}_j$ is the spin polarization direction of sublattice $j$ in the fixed layer, $g_e$ is the electron gyromagnetic ratio, $\mu_B$ is the Bohr magneton, $\eta$ is sublattice spin polarization (which is equal to the octupole polarization), $J$ is the applied current, $d$ is the thickness of the free layer, $\boldsymbol{m}_i = \boldsymbol{M}_i/M_s$ is the normalized magnetization vector for sublattice $i$, and $M_s$ is the sublattice saturation magnetization.

First, we show that such OTT terms result in a net vanishing torque on the sublattice in the case of equal intra- and inter-sublattice terms, i.e., $\beta_{ij} = \beta_1 = \beta_2$. Assuming a small misalignment angle, $\phi$, between the sublattice magnetizations of the fixed and free layer (Extended Data Fig. 5a), we have the following torques acting on the sublattice 3:

$$\begin{cases} \boldsymbol{\tau}_{31} = \frac{g_e\mu_B}{eM_sd}2\eta J \sin\left[\frac{2\pi}{3} + \phi\right]\boldsymbol{x} \,, \\ \boldsymbol{\tau}_{32} = \frac{g_e\mu_B}{eM_sd}2\eta J \sin\left[\frac{2\pi}{3} - \phi\right](-\boldsymbol{x}) \,, \\ \boldsymbol{\tau}_{33} = \frac{g_e\mu_B}{eM_sd}2\eta J \sin[\phi]\,\boldsymbol{x} \,. \end{cases} \tag{5}$$

We chose sublattice 3 for simplicity and the same analysis holds for all three sublattices. These torques are shown with red arrows in Extended Data Fig. 5a. The total torque, $\boldsymbol{\tau}_3 = \boldsymbol{\tau}_{33} + (\boldsymbol{\tau}_{31} + \boldsymbol{\tau}_{32})$, acting on sublattice 3 is given by

$$\boldsymbol{\tau}_3 = \frac{g_e\mu_B}{eM_sd}2\eta J\left\{\sin\left[\frac{2\pi}{3} + \phi\right] - \sin\left[\frac{2\pi}{3} - \phi\right] + \sin[\phi]\right\}\boldsymbol{x} = 0 \,. \tag{6}$$

Therefore, in the case of $\beta_1 = \beta_2$, there is no net torque on the sublattices as expected. For the case of $\beta_1 \neq \beta_2$, the net torque on sublattice 3 is given by

$$\boldsymbol{T}_3 = \beta_1 \boldsymbol{\tau}_{33} + \beta_2(\boldsymbol{\tau}_{31} + \boldsymbol{\tau}_{32})$$

$$= \frac{g_e \mu_B}{e M_s d} 2\eta(\beta_1 - \beta_2) J \sin[\phi]\, \boldsymbol{x}$$

$$= \frac{g_e \mu_B}{e M_s d} 2\eta \beta_{eff} J\, \boldsymbol{m}_3 \times (\boldsymbol{m}_3 \times \boldsymbol{p}_3). \quad (7)$$

This is a non-vanishing intra-sublattice torque with an effective OTT efficiency of $\beta_{eff} = \beta_1 - \beta_2$. Therefore, we use the following effective net intra-sublattice torques (demonstrated with red arrows in Extended Data Fig. 5a) for all three sublattices in our micromagnetic simulations:

$$\boldsymbol{T}_i = \frac{g_e \mu_B}{e M_s d} 2\eta \beta_{eff} J\, \boldsymbol{m}_i \times (\boldsymbol{m}_i \times \boldsymbol{p}_i). \quad (8)$$

To describe the magnetization dynamics of the free layer, we utilize a generalized three-sublattice model[65,66] in which the magnetization dynamics of the antiferromagnet are given by three coupled LLG equations for the sublattices:

$$\begin{cases} \frac{\partial \boldsymbol{m}_1}{\partial t} = -\frac{\mu_0}{(1+\alpha^2)}\left[(\boldsymbol{m}_1 \times \boldsymbol{H}_{\text{eff},1}) + \alpha \boldsymbol{m}_1 \times (\boldsymbol{m}_1 \times \boldsymbol{H}_{\text{eff},1}) + \boldsymbol{T}_1\right] \\ \frac{\partial \boldsymbol{m}_2}{\partial t} = -\frac{\mu_0}{(1+\alpha^2)}\left[(\boldsymbol{m}_2 \times \boldsymbol{H}_{\text{eff},2}) + \alpha \boldsymbol{m}_2 \times (\boldsymbol{m}_2 \times \boldsymbol{H}_{\text{eff},2}) + \boldsymbol{T}_2\right], \\ \frac{\partial \boldsymbol{m}_3}{\partial t} = -\frac{\mu_0}{(1+\alpha^2)}\left[(\boldsymbol{m}_3 \times \boldsymbol{H}_{\text{eff},3}) + \alpha \boldsymbol{m}_3 \times (\boldsymbol{m}_3 \times \boldsymbol{H}_{\text{eff},3}) + \boldsymbol{T}_3\right] \end{cases} \quad (9)$$

where $\alpha$ is the Gilbert damping. The effective field $\boldsymbol{H}_{\text{eff},i}$ is given by

$$\boldsymbol{H}_{\text{eff},i} = \sum_{i \neq j} H_{\text{exch}} \boldsymbol{m}_j + \boldsymbol{H}_{\text{ani},i}\,, \quad (10)$$

where the first term describes the antiferromagnetic exchange interaction between sublattice $i$ and sublattice $j$, where $H_{\text{exch}} = 4A_{AB}/(M_{s,i} a_0^2)$ is the strength of the field, with $A_{AB}$ the inter-sublattice exchange coupling and $a_0$ the lattice constant. The second term is the anisotropy field for sublattice $i$ along the easy-axis direction $\boldsymbol{e}_i$, given by

$$\boldsymbol{H}_{\text{ani},i} = 2\frac{K_{\text{u}}}{M_{s,i}} \boldsymbol{m}_i \cdot \boldsymbol{e}_i\,. \quad (11)$$

The easy axes are given by $\boldsymbol{e}_1 = \left(-\frac{\sqrt{3}}{2}, -\frac{1}{2}, 0\right)$, $\boldsymbol{e}_2 = \left(+\frac{\sqrt{3}}{2}, -\frac{1}{2}, 0\right)$ and $\boldsymbol{e}_3 = (0,1,0)$ as indicated in Extended Data Fig. 5a. We initialize the sublattice magnetizations along directions $-\boldsymbol{e}_1$, $-\boldsymbol{e}_2$, $-\boldsymbol{e}_3$, respectively, and inject a current density $J = 10.0$ MA/cm$^2$ into the stack. Extended Data Fig. 5b and c summarize the switching process considering the parameters listed in Extended Data Table 1. Extended Data Fig. 5b (c) presents the time-evolution of the $x$ ($y$) component of the magnetization of each sublattice for a direct (reverse) current with solid (dashed) lines. The plots show that the sublattice magnetizations remain at 120$^\circ$ throughout the dynamics because of the large exchange energy. Such a dynamic represents the octupole moment switching. This can be understood as the magnitude of the net torque on the sublattices is equal and their direction is related by 120$^\circ$ rotation, as indicated in Extended Data Fig. 5a with the green arrows. The collective effect of these torques can thus be considered as an effective octupole spin-transfer torque acting on the octupole moment of the free layer.

**Extended Data Table 1|** The three-sublattice model parameters utilized for the simulation shown in Extended Data Fig. 5.

| **Parameter** | **Value** | **Unit** |
|---|---|---|
| Saturation magnetization, $M_s$ | 153.0 | kA/m |
| Uniaxial anisotropy energy density, $K_u$ | 10.0 | kJ/m$^3$ |
| Exchange energy, $A_{AB}$ | -10.0 | pJ/m |
| Gilbert damping, $\alpha$ | 0.01 | |
| Sublattice spin polarization, $\eta$ | 0.53 | |
| Effective OTT efficiency, $\beta_{eff}$ | 1 | |
| Lattice constant, $a_0$ | 0.5 | nm |
| FL thickness, $d$ | 10.0 | nm |

**Comparison with epitaxial $PtMn_3$ (111) films**

The textured $PtMn_3$ (111) films and corresponding full AATJ stacks did not show a measurable uncompensated magnetic moment, and could not be switched by a magnetic field within our experimentally accessible range of fields. Supplementary Note 7 compares these films to epitaxially grown $PtMn_3$ (111) films, which do exhibit an uncompensated magnetic moment as evidenced in SQUID magnetometry measurements, as well as an anomalous Hall signal that can be switched by the applied magnetic field. The immunity to external fields is further shown in Supplementary Note 8, in which an AATJ exhibits similar two-terminal switching behavior under large external magnetic fields, without significant changes in switching current values or TMR.

## Data availability

The data that support the plots in this paper and other findings of this study are available from the corresponding authors upon request. The scientific color maps developed by Crameri et al.[67] are used in this study to prevent visual distortion of the data and exclusion of readers with colour vision deficiencies[68].

51. Szunyogh, L., Lazarovits, B., Udvardi, L., Jackson, J. & Nowak, U. Giant magnetic anisotropy of the bulk antiferromagnets IrMn and $IrMn_3$ from first principles. *Phys. Rev. B* **79**, 020403(R) (2009).

52. Yamane, Y., Gomonay, O. & Sinova, J. Dynamics of noncollinear antiferromagnetic textures driven by spin current injection. *Phys. Rev. B* **100**, 054415 (2019).

53. Shukla, A. & Rakheja, S. Spin-torque-driven terahertz auto-oscillations in noncollinear coplanar antiferromagnets. *Phys. Rev. Appl.* **17**, 034037 (2022).

54. Ulloa, C. & Nunez, A. S. Solitonlike magnetization textures in noncollinear antiferromagnets. *Phys. Rev. B* **93**, 134429 (2016).

55. Dasgupta, S. & Tchernyshyov, O. Theory of spin waves in a hexagonal antiferromagnet. *Phys. Rev. B* **102**, 144417 (2020).
56. He, Z. & Liu, L. Magnetic dynamics of strained non-collinear antiferromagnet. *J. Appl. Phys.* **135**, 093902 (2024).
57. Nelapudi, L. S., Konakanchi, S. T. & Upadhyaya, P. *SPICE-based compact model for chiral antiferromagnet system with spin current injection* (Version 1.0.0). nanoHUB https://doi.org/10.21981/W9Z6-YV76 (2025).
58. Hu, S., Zheng, C., Chen, C., Zhou, Y. & Liu, Y. Current-driven spin oscillations in noncollinear antiferromagnetic tunnel junctions. *Phys. Rev. B* **109**, 174433 (2024).
59. Krén, E. et al. Magnetic structures and exchange interactions in the Mn-Pt system. *Phys. Rev.* **171**, 574-585 (1968).
60. Kresse, G. & Furthmüller, J. Efficient iterative schemes for ab initio total-energy calculations using a plane-wave basis set. *Phys. Rev. B* **54**, 11169 (1996).
61. Kresse, G. & Joubert, D. From ultrasoft pseudopotentials to the projector augmented-wave method. *Phys. Rev. B* **59**, 1758 (1999).
62. Perdew, J. P. et al. Restoring the density-gradient expansion for exchange in solids and surfaces. *Phys. Rev. Lett.* **100**, 136406 (2009).
63. Liechtenstein, A. I., Anisimov, V. I. & Zaanen, J. Density-functional theory and strong interactions: Orbital ordering in Mott-Hubbard insulators. *Phys. Rev. B* **52**, R5467(R) (1995)
64. Mostofi, A. A. et al. An updated version of wannier90: A tool for obtaining maximally-localised Wannier functions. *Comput. Phys. Commun.* **185**, 239 (2014)
65. Tomasello, R. et al. Domain periodicity in an easy-plane antiferromagnet with Dzyaloshinskii-Moriya interaction. *Phys. Rev. B* **102**, 224432 (2020).
66. Tomasello, R. et al. Antiferromagnetic Parametric Resonance Driven by Voltage-Controlled Magnetic Anisotropy. *Phys. Rev. Appl.* **17**, 034004 (2022).
67. Crameri, F. *Scientific Colour Maps*, https://zenodo.org/records/5501399 (2021).
68. Crameri, F., Shephard, G. E. & Heron, P. J. The misuse of colour in science communication. *Nat. Commun.* **11**, 5444 (2020).

## Acknowledgements

This research was primarily supported as part of the Center for Energy-Efficient Magnonics (CEEMag), an Energy Frontier Research Center funded by the U.S. Department of Energy (DOE), Office of Science, Basic Energy Sciences (BES), under Award number DE-AC02-76SF00515 (design of experiments, electrical measurements, material growth, NV microscopy, theoretical analysis, and data analysis) and by the U.S. National Science Foundation (NSF) under Award numbers CNS-2425538 (pulse width dependence and energy barrier measurement at Northwestern University, electrical measurements at UT Austin), DMR-2308691 (TEM analysis and charge transport characterization at Northwestern University), and DMR-2425567 (theoretical modeling at the University of Nebraska-Lincoln). We acknowledge useful discussions with Dr. Soho Shim on scanning NV microscopy and with Dr. Charudatta Phatak on TEM analysis. MH thanks the Swiss National Science Foundation (SNSF) for the financial support via SNSF Postdoc.Mobility Grant No. 222234 (theoretical analysis). LDY and JMR acknowledge support by the Air Force Office of Scientific Research under award number FA9550-23-1-0042 (DFT simulations, theoretical analysis). W.R. was supported by the SUPeRior Energy-efficient Materials and dEvices (SUPREME) Center, one of seven centers in the JUMP 2.0, a Semiconductor Research Corporation (SRC) program sponsored by DARPA (Wannier models, theoretical analysis). AM, MC and GF acknowledge support by the Contract n. 2025-40-I.0 (SPINAM) funded by the Italian Space Agency within the call “Studi di concetti innovativi di sistemi spaziali” (micromagnetic simulations). In addition, this work made use of the EPIC facility (RRID: SCR_026361) of Northwestern University’s NUANCE Center, which has received support from the SHyNE Resource (NSF ECCS-2025633), the International Institute of Nanotechnology (IIN) (NIH-S10OD026871), and Northwestern's MRSEC program (NSF DMR-2308691).

## Author contributions

PKA, JK and MH conceived the idea. PKA designed the AATJ material stack and devices. LW, SCM, and JAK fabricated the devices. JK, SKC, MSNT and MH carried out electrical measurements with support from JA, SA, MCH, EM, AD, JF, and VKS. MH and JK designed the NV magnetometry sample. MH and EM grew and characterized the NV magnetometry sample with support from JK. AD'A and GDF performed NV center microscopy. PGL performed the TEM analysis with support from JK, MH, VPD and MCH. MH, JK, and EM performed the XRD and SQUID magnetometry measurements. NZ, JACI, JK, and SKC performed the comparison of two- and three-terminal switching geometries. MH developed the theoretical description of OTT. MH, LDY, ME, WR, EYT, and JMR performed the DFT calculations and theoretical analysis. AM, MC and GF developed the micromagnetic code to simulate the three sub-lattice model. AM, MC, and GF performed the micromagnetic simulations with input from MH and PKA. STK and PU performed the theoretical energy barrier calculation. JK, MH and PKA wrote the manuscript with contributions from all the other authors. All authors discussed the results, contributed to the data analysis, and commented on the manuscript. The study was performed under the leadership of PKA.

## Competing interests

The authors declare no competing interests.

## Additional information

**Correspondence and requests for materials** should be addressed to Jaimin Kang, Mohammad Hamdi or Pedram Khalili Amiri.

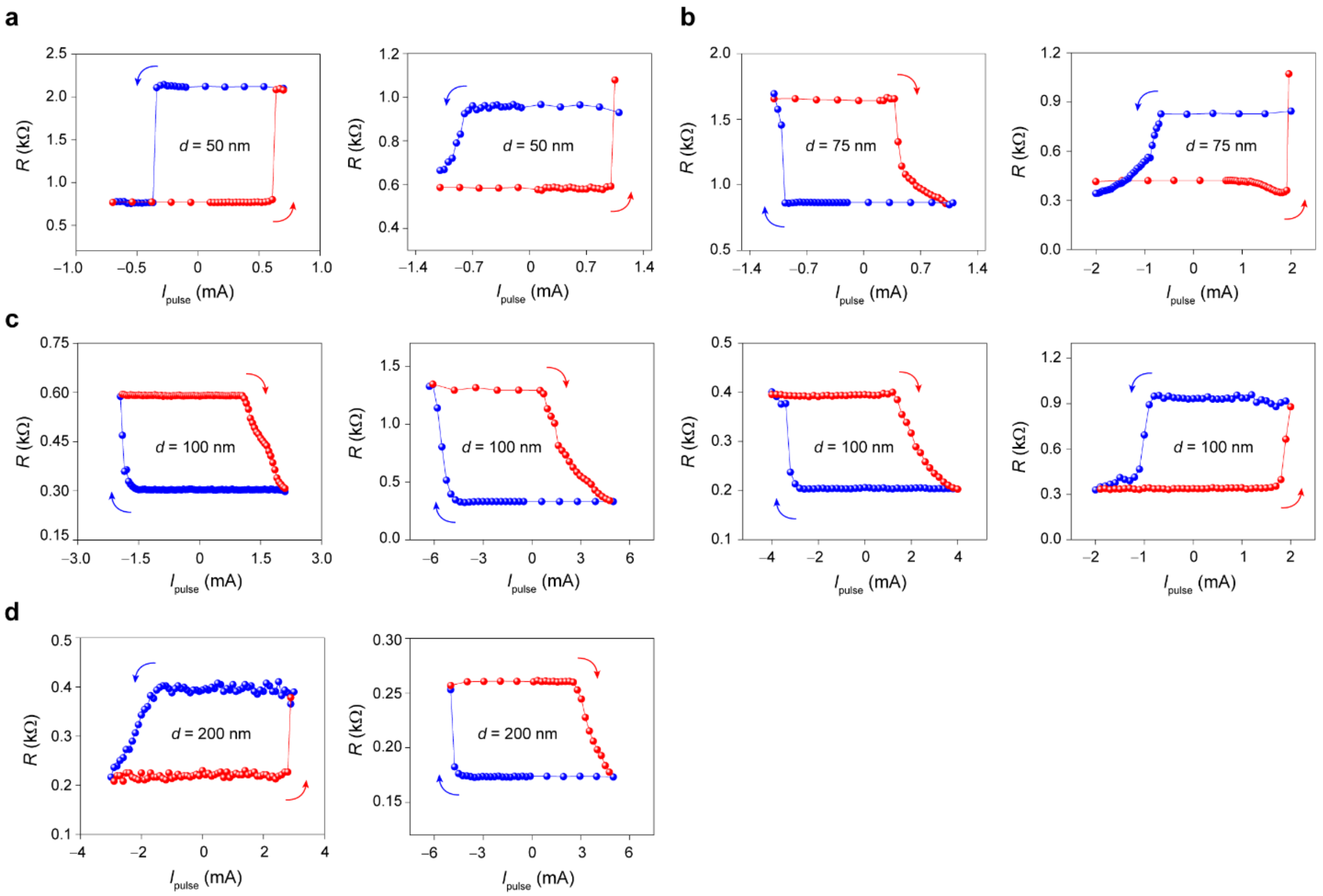

**Extended Data Fig. 1| Two-terminal current-induced switching of AATJs with different diameters of pillars. a-d,** Representative current-induced switching measurements performed with the different diameters of 50 nm (**a**), 75 nm (**b**), 100 nm (**c**) and 200 nm (**d**).

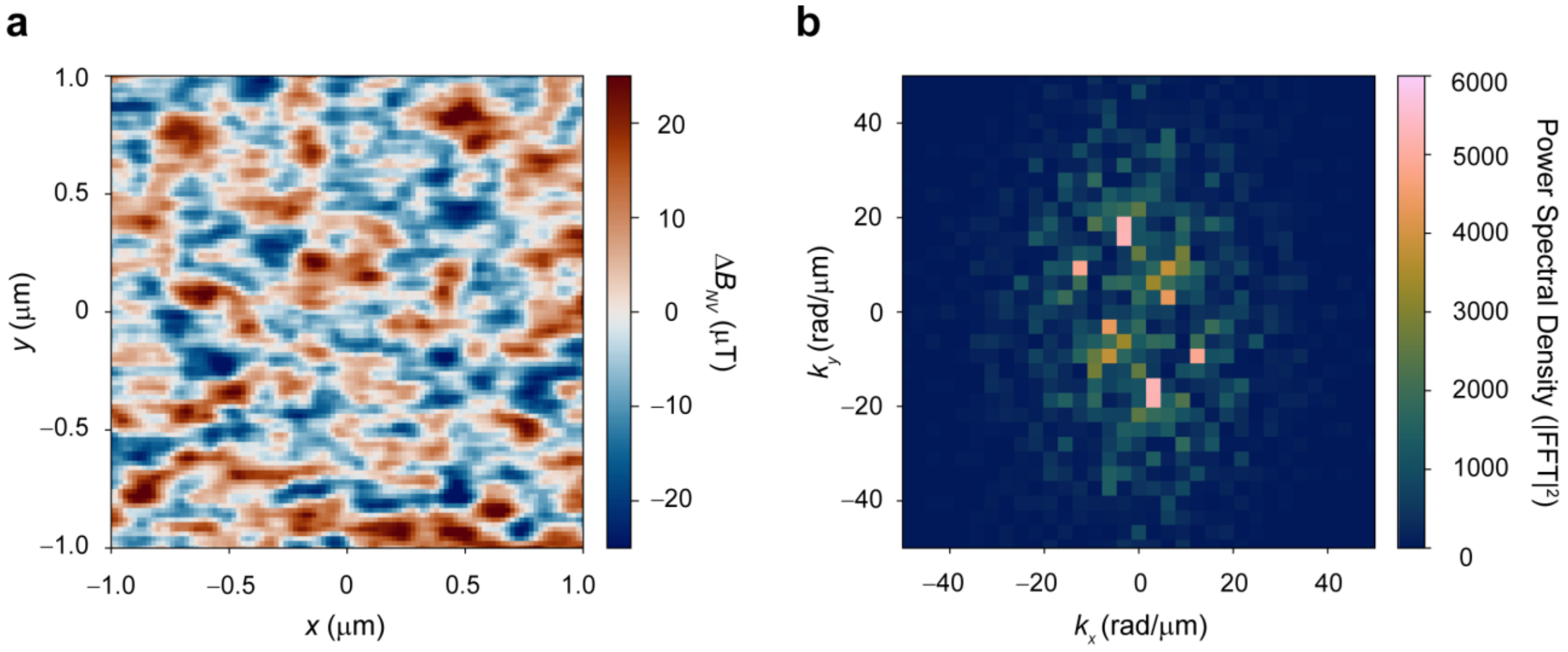

**Extended Data Fig. 2| NV magnetometry.** **a**, Scanning NV magnetometry measurements performed 50 nm above a MgO (100) / $PtMn_3$ (100) / Pt (5 nm) sample. **b**, Corresponding Fourier transform of the NV magnetometry measurements.

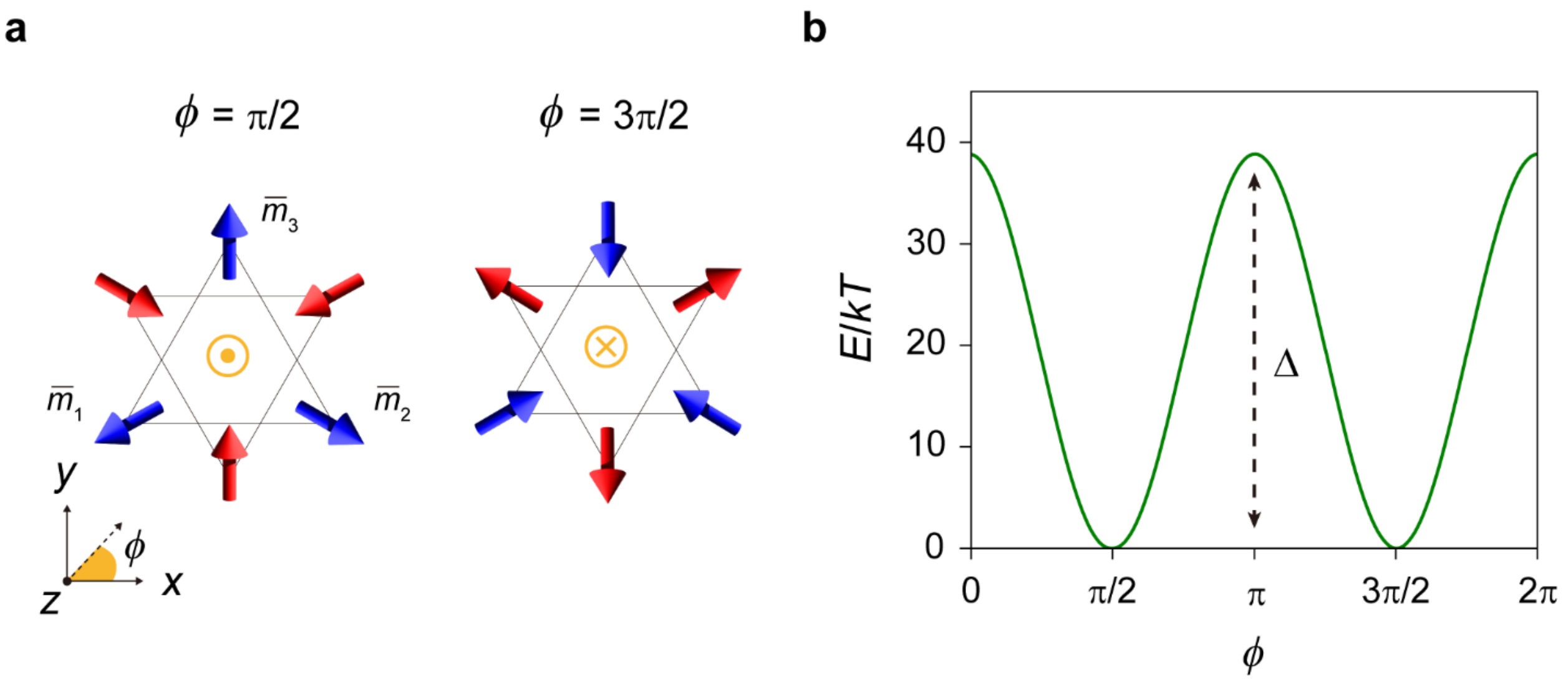

**Extended Data Fig. 3| Calculation of energy barrier of the $PtMn_3$ free layer. a**, Schematic of the unit cell of $PtMn_3$ viewed along [111] showing two equilibrium configurations. The Kagome plane is the *x*-*y* plane. **b**, Energy landscape of a 100 nm wide and 7 nm thick nanodot as a function of the rigid rotation of the $PtMn_3$ spin motif from panel a, depicting a two-fold ground state with an energy barrier of $\Delta \approx 39$ kT.

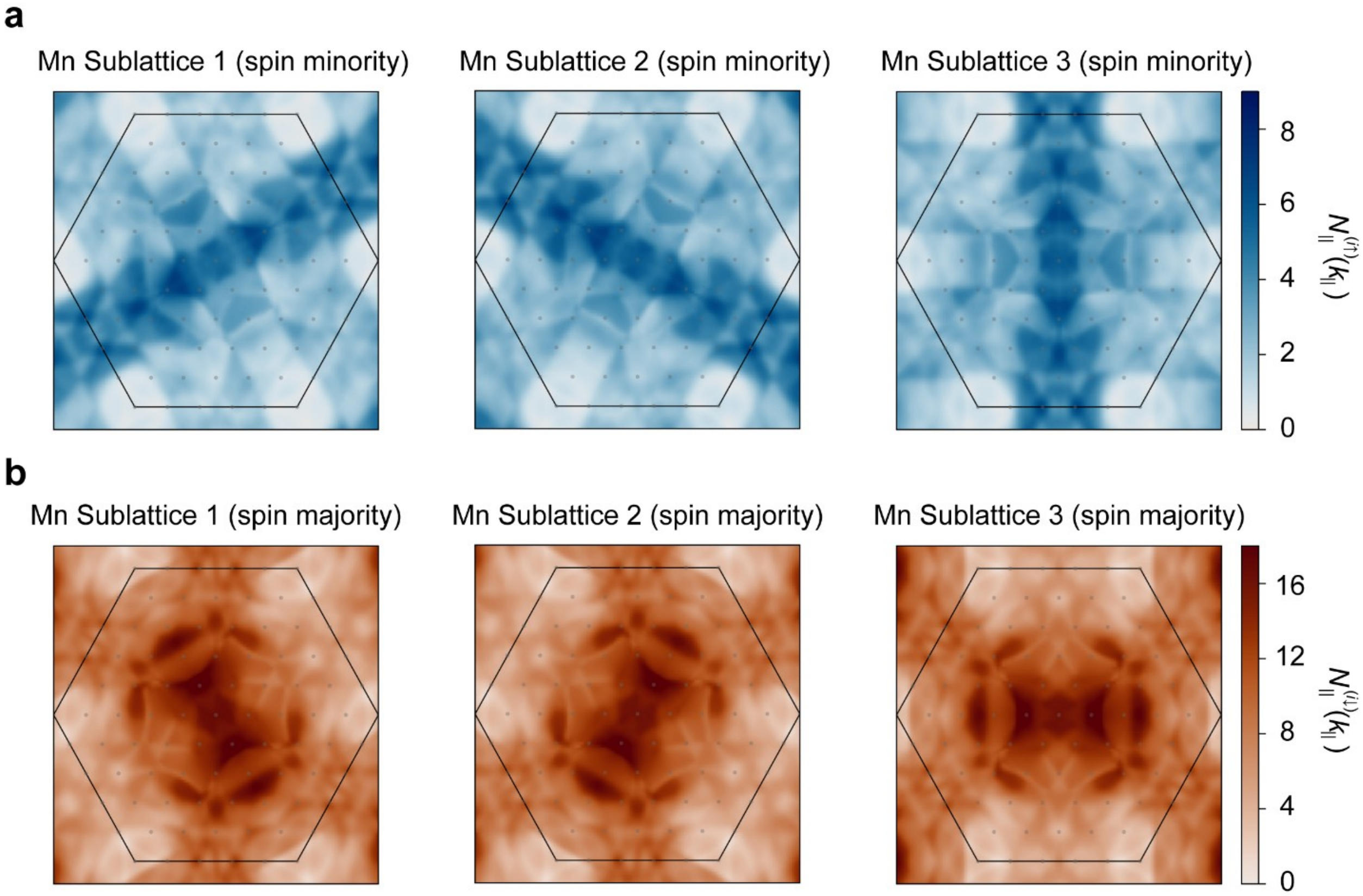

**Extended Data Fig. 4| Sublattice site-spin projected conduction channels as a function of $k_{||}$ for $PtMn_3$. a**, **b**, Conduction channels, projected onto the site-spin direction of the corresponding sublattice for the spin minority **(a)** and spin majority **(b)** channels.

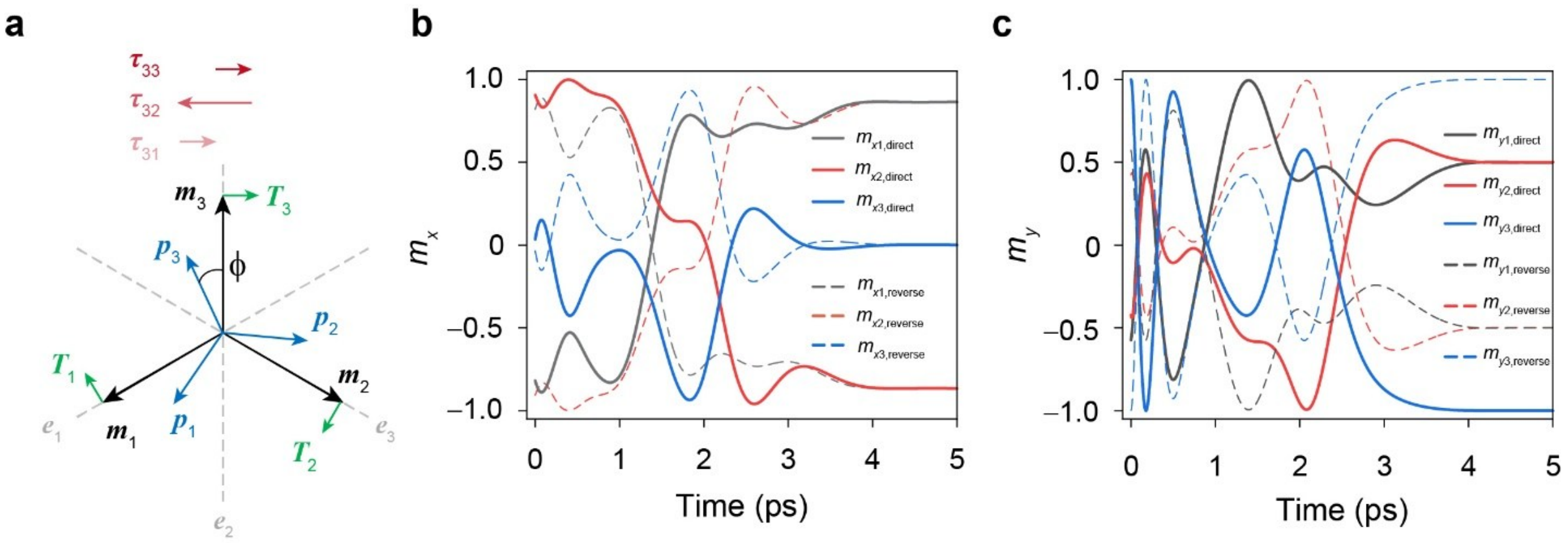

**Extended Data Fig. 5| Macrospin simulation of OTT switching in AATJ. a**, Schematic of the simulated three-sublattice system. Black, blue, red and green arrows indicate the sublattice magnetization in the free layer, spin polarization of the sublattices in the fixed layer, intra- and inter-sublattice torques acting on sublattice 3 in the free layer for $\beta_1 = \beta_2$, and total torques acting on the respective sublattices for $\beta_1 \neq \beta_2$. Gray dashed lines indicate the respective easy axis of a given sublattice. $\phi$ is a small misalignment angle (exaggerated here) between the respective sublattice magnetizations in the free layer and sublattice spin polarizations in the fixed layer. **b**, **c**, Time-dependent *x*-component **(b)** and *y*-component **(c)** of the three sublattice magnetizations. Solid (dashed) lines represent the switching via a direct (reverse) current from AP to P (P to AP) states.

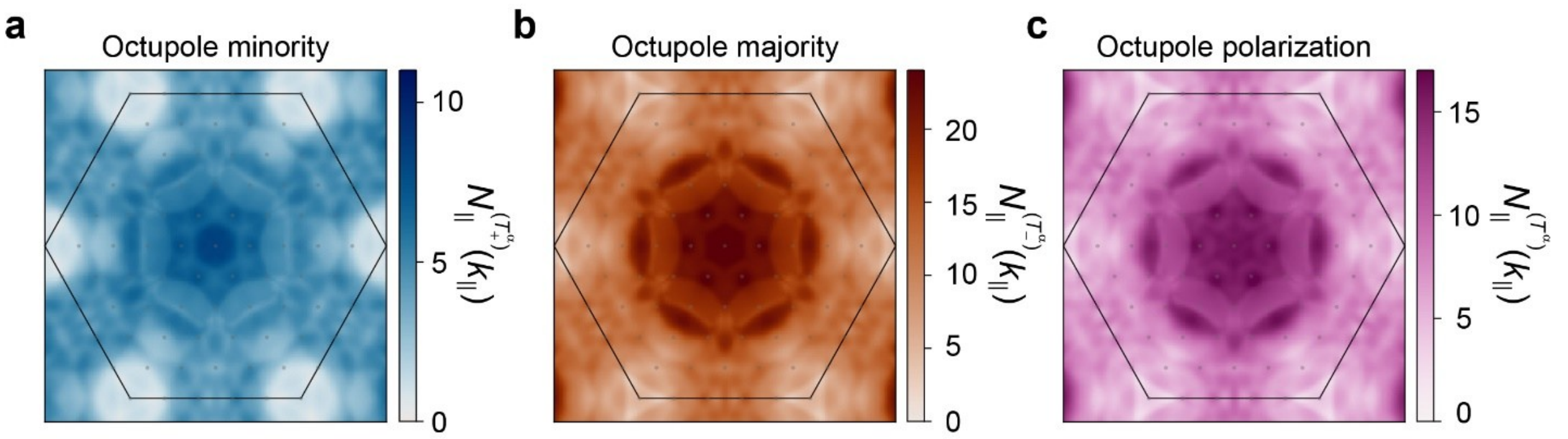

**Extended Data Fig. 6| Octupole-projected conduction channels as a function of $k_{||}$ for $PtMn_3$.** **a**, Octupole minority, **b**, octupole majority and, **c**, the net octupole polarization-resolved conduction channels.

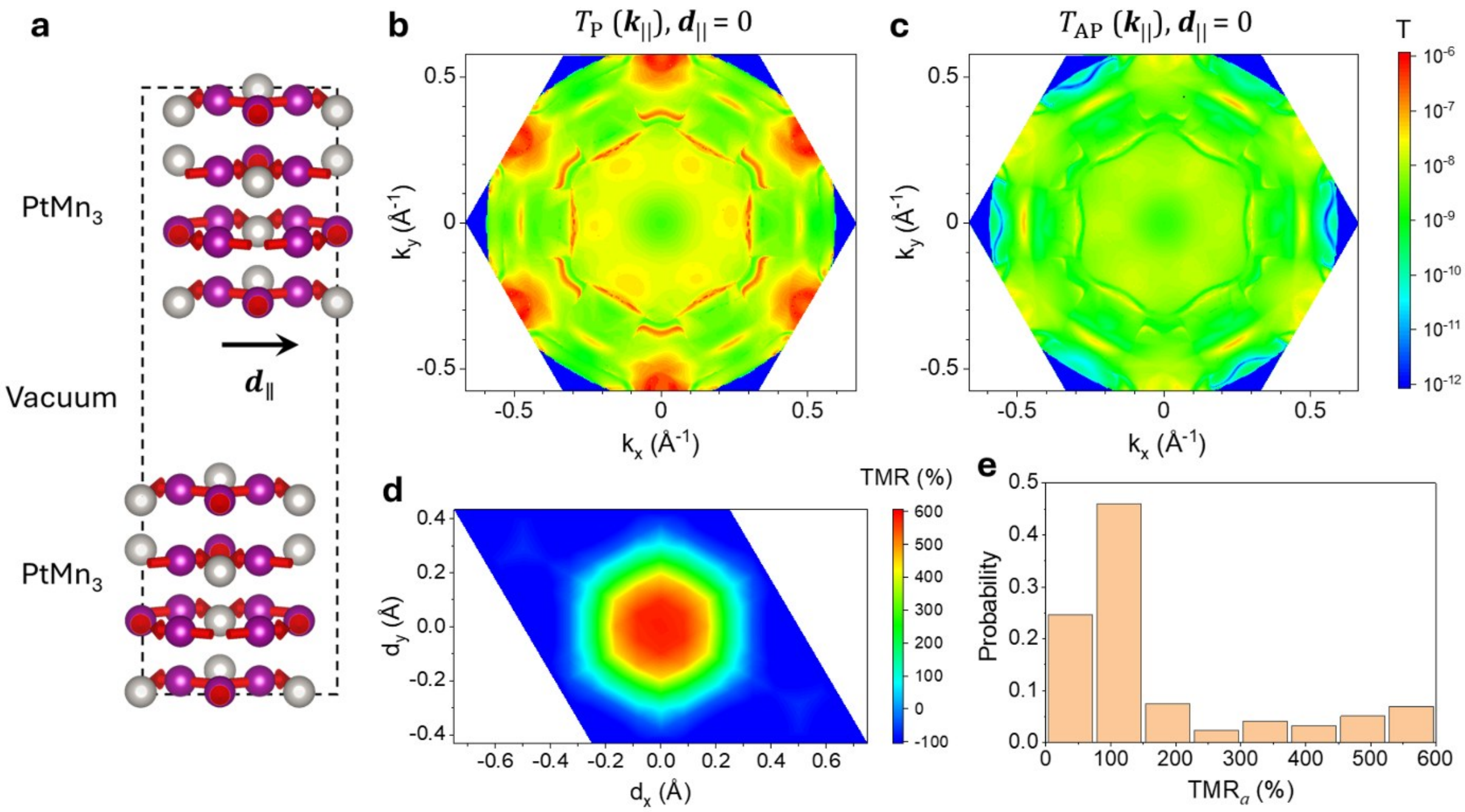

**Extended Data Fig. 7| Theoretical modelling of the effect of interface termination on TMR in PtMn3/vacuum/PtMn3 AATJs**. **a**, Schematic of an AATJ where the top PtMn3 layer has a lateral shift of $\boldsymbol{d}_{\parallel}$ with respect to the bottom layer. **b**, **c**, $\boldsymbol{k}_{\parallel}$- resolved transmission for parallel $T_P(\boldsymbol{k}_{\parallel})$ (**b**) and antiparallel $T_{AP}(\boldsymbol{k}_{\parallel})$ (**c**) alignment of antiferromagnetic PtMn3 layers for perfect matching of the magnetic interface structure for top and bottom electrodes ($\boldsymbol{d}_{\parallel} = 0$). **d,** $TMR = (T_P - T_{AP})/T_{AP}$ as a function of lateral shift $\boldsymbol{d}_{\parallel} = (d_x, d_y)$. **e**, Histogram of the distribution of TMR ratios defined as $TMR_a = |T_P - T_{AP}|/\min(T_{P,}T_{AP})$.

**Supplementary Note 1. Structural and magnetic characterization of $PtMn_3$ films**

To determine the structural and magnetic properties of our $PtMn_3$ films, we performed X-ray diffraction (XRD) and superconducting quantum interference device (SQUID) magnetometry measurements on the AATJ stack. In the XRD data, shown in Fig. S1, only the Pt (111) and $PtMn_3$ (111) peaks are observed. This is consistent with the highly textured films observed in the TEM measurements (main text, Fig. 1d) and does not show any evidence for the existence of other Pt-Mn phases. SQUID magnetometry data on the full AATJ stack are shown in Fig. S2, where the magnetic field is swept along both the in-plane and out-of-plane directions up to a magnitude of 7 T. The data show no net magnetization in either direction, indicating the fully compensated antiferromagnetic phase of $PtMn_3$. In addition, to determine the actual stoichiometry of the $PtMn_3$ films, X-ray fluorescence (XRF) measurements were performed. The Pt composition was found to range from 23.9 to 25.8 atomic percent (at%) at multiple positions across the 150 mm silicon wafer, which is close to the targeted $Pt_{25}Mn_{75}$ composition. These results are illustrated in Fig. S3.

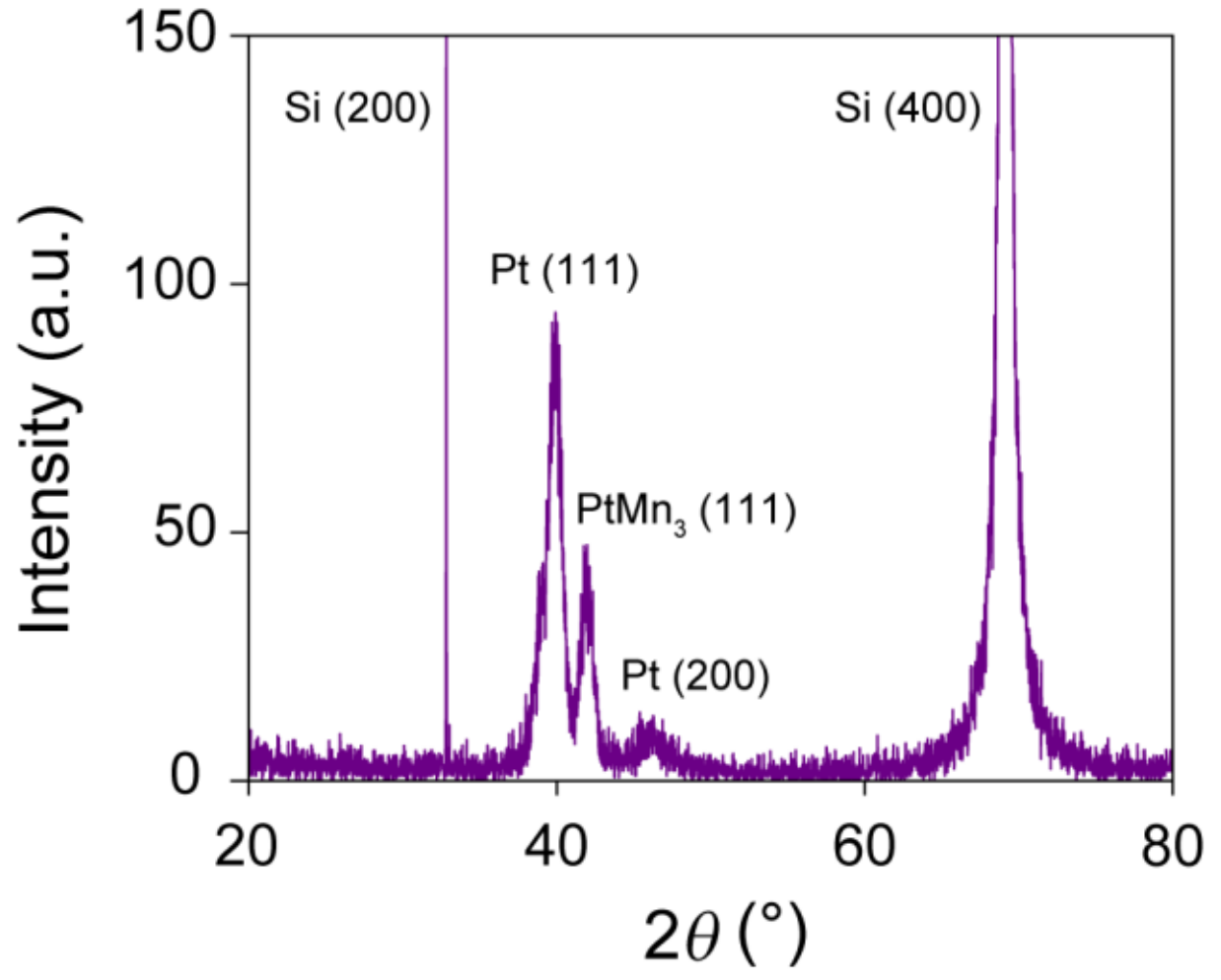

**Figure S1| X-ray diffraction measurements.** XRD measurements on full Si / $SiO_2$ / Pt (5 nm) / $PtMn_3$ (7 nm) / MgO (2 nm) / $PtMn_3$ (10 nm) / Pt (5 nm) / Ru (10 nm) AATJ material stack.

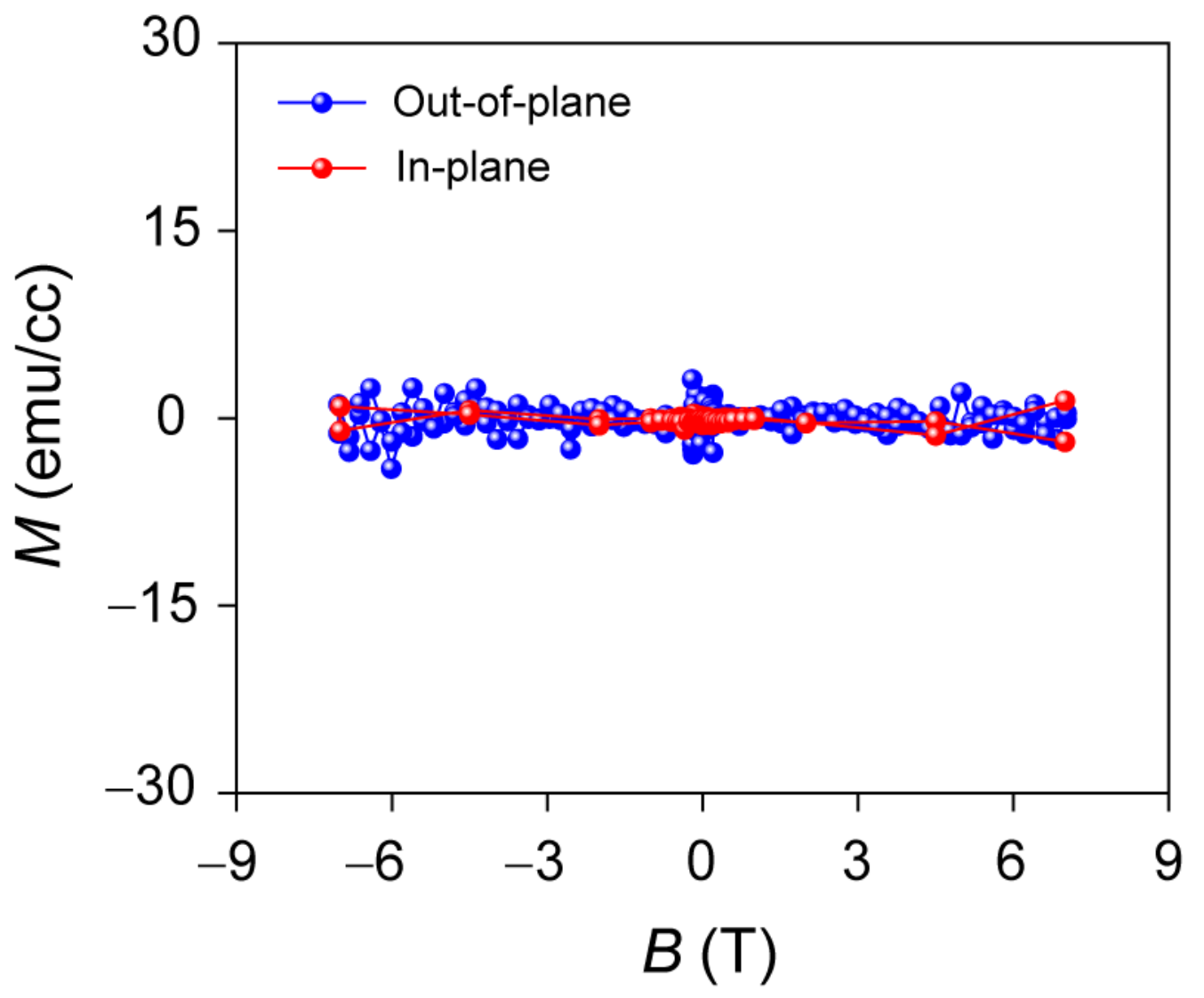

**Figure S2| SQUID magnetometry measurements.** Magnetization measurements on full Si / $SiO_2$ / Pt (5 nm)/ $PtMn_3$ (7 nm) / MgO (2 nm) / $PtMn_3$ (10 nm) / Pt (5 nm) / Ru (10 nm) AATJ material stack, conducted using a superconducting quantum interference device by sweeping the external magnetic field along the in-plane (red) and out-of-plane (blue) directions. The curves were obtained after subtracting the diamagnetic linear background.

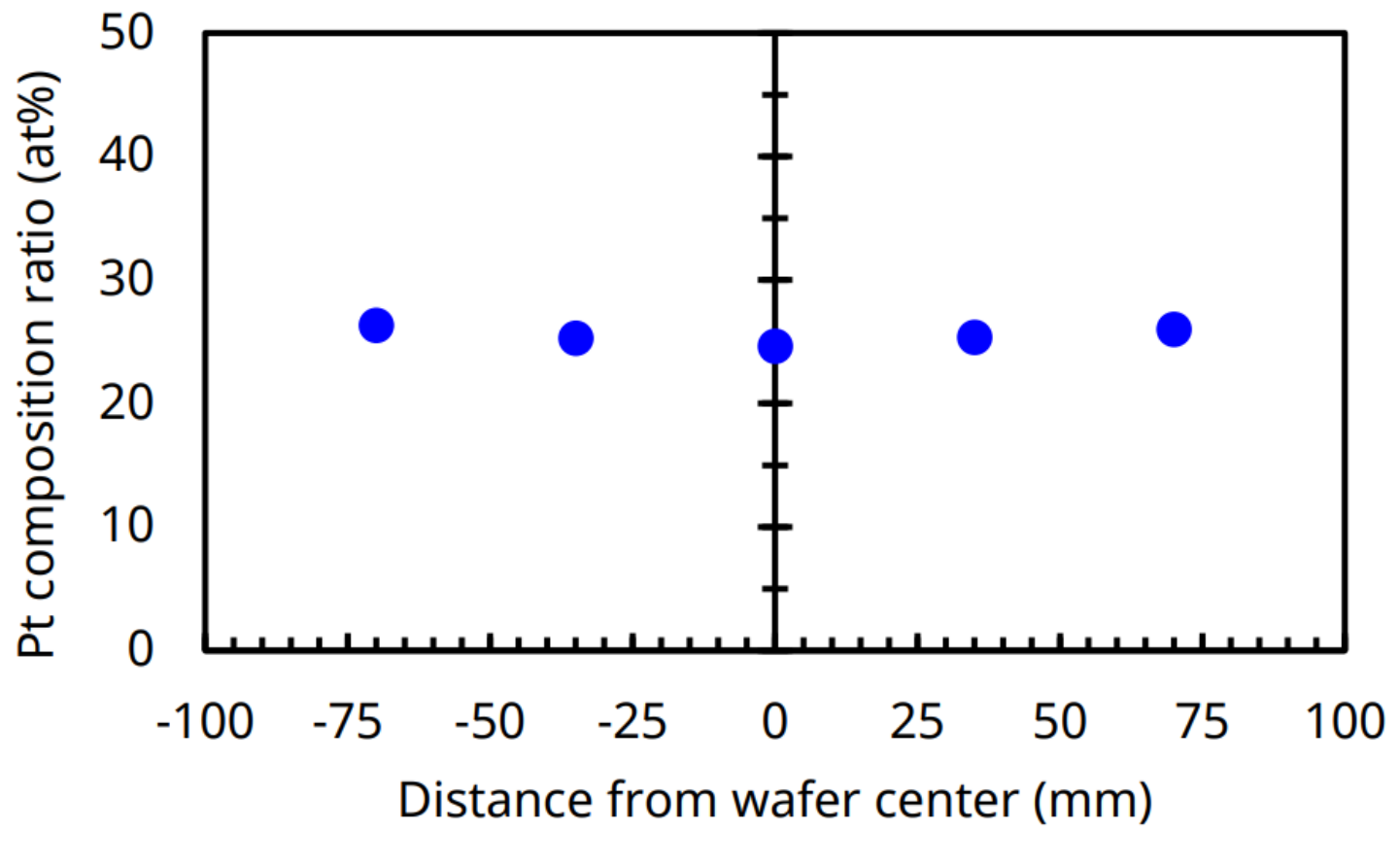

**Figure S3| Composition of $PtMn_3$ across different positions on a 150 mm wafer.** The composition was measured by X-ray fluorescence, indicating a uniform stoichiometry.

**Supplementary Note 2. Comparison of $PtMn_3$ (001) and (111) surface terminations**

To estimate the intrinsic domain size in sputter-deposited $PtMn_3$ using NV magnetometry, we used a $PtMn_3$ (001) film to obtain an uncompensated surface termination with finite stray field. Using the $PtMn_3$ (111) surface for this purpose was not possible, since it is fully compensated. A comparison of these two surfaces of $PtMn_3$ is shown in Fig. S4.

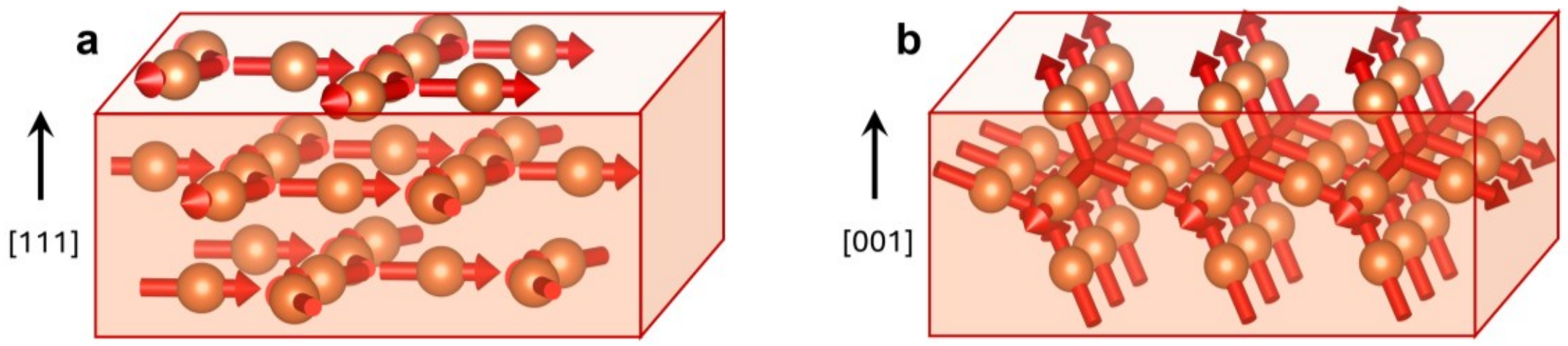

**Figure S4| Comparison of (001) and (111) surface terminations of $PtMn_3$.** Only the (001) surface has an uncompensated magnetic moment, which allows its domain structure to be detected in NV microscopy measurements.

**Supplementary Note 3. Role of amorphous MgO in TMR of AATJ devices**

In crystalline MgO, the barrier is translationally symmetric in-plane, so tunneling is specular, resulting in the conservation of the transverse electron momentum $\boldsymbol{k}_{\parallel}$. In amorphous MgO, this symmetry is lost. Random fluctuations in atomic positions create local variations of the barrier height experienced by tunneling electrons, generating a scattering potential within the barrier. As a result, tunneling becomes diffusive and the transverse momentum is no longer conserved, $\boldsymbol{k}_{\parallel} \neq \boldsymbol{k}'_{\parallel}$. Typically, however, these fluctuations are small compared to the average barrier height, $\Delta \ll V_0$. Specifically, for Fe/MgO/Fe tunnel junctions, the effective barrier height is typically $V_0 \sim$ 2–3 eV, as inferred from ab-initio band alignment and experimental RA–vs–thickness fits for Fe/MgO/Fe MTJs. The same magnitude of $V_0$ is expected in $PtMn_3$/MgO/$PtMn_3$ junctions. On the other hand, the intrinsic fluctuations of the conduction band minimum (CBM), $\Delta$, in an amorphous oxide layer, as follows from typical magnitudes of band-edge Urbach tail energies in amorphous oxides are of the order of $\Delta \sim$ 0.3–0.4 eV [1]. Thus, $\frac{\Delta}{V_0} \sim$ 0.1–0.2 , can be considered as a small perturbative parameter.

This physically relevant limit justifies using the Born approximation: fluctuations of the barrier height perturb the evanescent MgO barrier only weakly, allowing tunneling to be treated as a small single-scattering correction on top of the clean barrier. In this case, transmission can then be written as a sum of a coherent and a diffusive part:

$$T(\boldsymbol{k} \to \boldsymbol{k}') = T_0(\boldsymbol{k}_{\parallel})\, \delta_{\boldsymbol{k}_{\parallel},\boldsymbol{k}'_{\parallel}} + \delta T_{\mathrm{diff}}(\boldsymbol{k} \to \boldsymbol{k}').$$

The first term is the usual specular, $\boldsymbol{k}_{\parallel}$-conserving tunneling through a clean MgO barrier. The second term describes $k$-mixing due to disorder. For Gaussian-correlated barrier-height fluctuations with correlation length, $\xi$, the diffusive contribution can be shown to have the form

$$\delta T_{\text{diff}}(\boldsymbol{k} \to \boldsymbol{k}') \propto \Delta^2 \xi^3 L\, e^{-\frac{\xi^2}{2}|\boldsymbol{k}'_\parallel - \boldsymbol{k}_\parallel|^2},$$

where $\Delta$ is the RMS amplitude of barrier-height fluctuations, $\xi$ is their correlation length, and $L$ is the barrier thickness.

For an amorphous MgO oxide, it is reasonable to take $\xi$ to be of the order of half a lattice parameter, i.e., $\xi \sim 2$ Å. Then the *k*-space width of the Gaussian is $\sim \frac{1}{\xi} \approx 0.5\ \text{Å}^{-1}$, which is a substantial fraction of the in-plane Brillouin-zone. In other words, an incoming $\boldsymbol{k}_\parallel$ is scattered into a fairly broad range of $\boldsymbol{k}'_\parallel$.

However, the integrated weight of this broadened, diffuse contribution is controlled by the small ratio $\Delta/V_0$. Within the Born approximation, integration of $\delta T(\boldsymbol{k} \to \boldsymbol{k}')$ over all final $\boldsymbol{k}'_\parallel$, the diffuse weight scales as

$$\frac{\delta T_{\text{diff}}}{T_0} \sim \kappa_0^2 L \xi \left(\frac{\Delta}{V_0}\right)^2,$$

where $\kappa_0$ is the decay rate in MgO, which is $\kappa_0 \sim 0.3\ \text{Å}^{-1}$. Now, taking into account that $\left(\frac{\Delta}{V_0}\right)^2 \sim 0.01\text{–}0.04$ and $L \sim 20$ Å, we obtain:

$$\frac{\delta T_{\text{diff}}}{T_0} \sim 0.03 - 0.12.$$

We see therefore that only a few percent of the total transmitted weight reside in the diffusive transmission, while the dominant part remains coherent and nearly specular.

**Supplementary Note 4. Temperature-dependent switching measurements for additional AATJ device**

For all tested devices, there was a complete disappearance of TMR and OTT switching at temperatures ranging from 360 K to 400 K. This process is reversible, i.e., the OTT-induced switching and TMR reappear upon return to room temperature, when the material goes back to its noncollinear order. Figure S5 shows a series of switching curves measured on an AATJ device starting from 300 K (blue circles), gradually increasing to 400 K (red circles) and back to 300 K (green circles). The results show that OTT switching behavior and associated TMR are absent at 360 K and 400 K, and reappear upon return to 300 K. We note that slight changes in the switching current, resistance values and associated TMR were observed in the recovered curve at 300 K. We attribute this to different local domain configurations formed during the heating and subsequent cooling process, as well as to a potential partial shorting of the device during the high-temperature measurements.

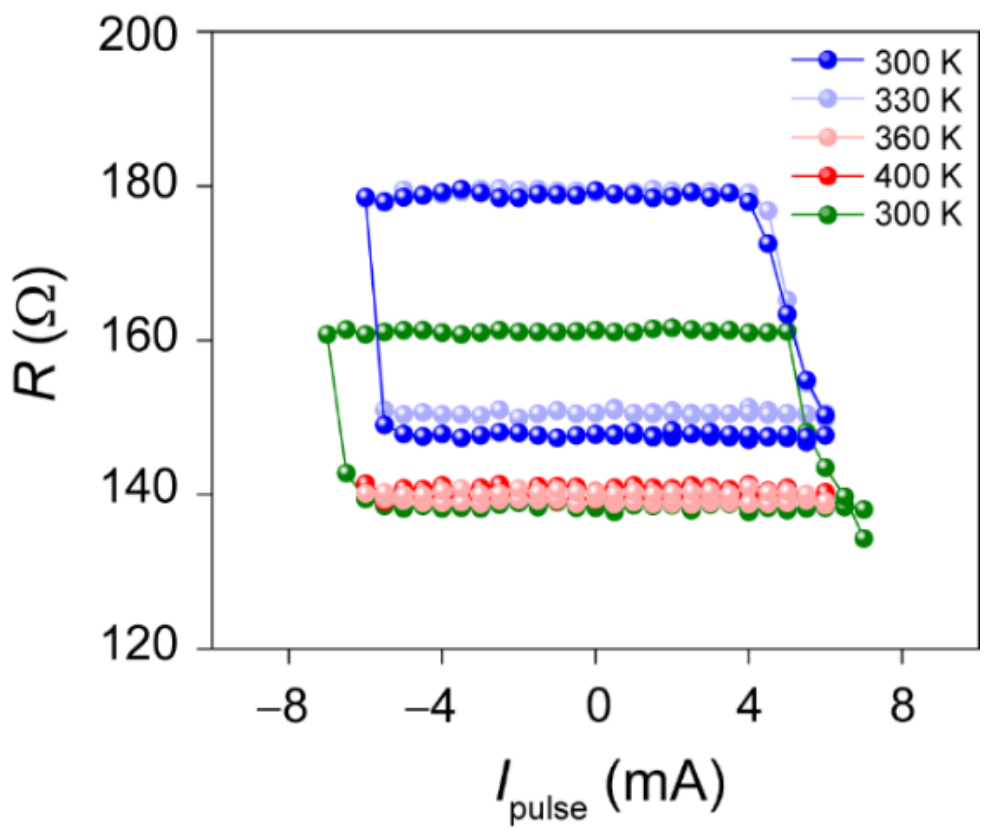

**Figure S5| Temperature-dependent current-induced switching measurements.** The switching measurements were performed for a temperature sequence starting from 300 K to 400 K and back to 300 K. Current-induced switching reappears upon cooling to room temperature, consistent with the octupolar origin of the OTT and TMR.

**Supplementary Note 5. Comparison of two- and three-terminal switching measurements**

We performed additional experiments to verify that spin-orbit torque (SOT) from the Pt seed layer is not responsible for the current-induced switching observed in our devices. Figure S6a shows a representative scanning electron microscope (SEM) image of the AATJ devices used for this experiment, prior to the top electrode deposition. The AATJ pillar is located at the center of the narrow Pt channel whose width is slightly larger than the diameter of the pillar. This design allows one to directly compare OTT and SOT switching in the same device, under an identical current density flowing through the Pt layer. Figure S6b shows the current-induced switching experiments in the OTT geometry (two-terminal, with write current flowing vertically through the AATJ) and the SOT geometry (three-terminal, with write current flowing horizontally through the Pt channel). Both OTT and SOT traces are measured in the same device with a pillar diameter of 100 nm. In both cases, the tunneling resistance was measured with a small reading current after each current-induced switching attempt. The results show OTT switching at a current pulse amplitude of approximately 2 mA, whereas switching is absent up to 4 mA of in-plane current in the SOT geometry. These results indicate that the observed switching signatures in the two-terminal OTT geometry do not originate from SOT.

Furthermore, we performed SOT and OTT switching experiments in a second device, after field-cooling the AATJ from 400 K to 300 K under the out-of-plane external magnetic field of ±0.2 T. The results of this experiment are summarized in Fig. S7. First, we do not observe SOT switching within the applied in-plane current range, consistent with the data shown in Fig. S6. Second, the polarity of the two-terminal OTT switching curves obtained under different field-cooling conditions remains the same, as would be expected from a spin-transfer torque-based switching mechanism [2].

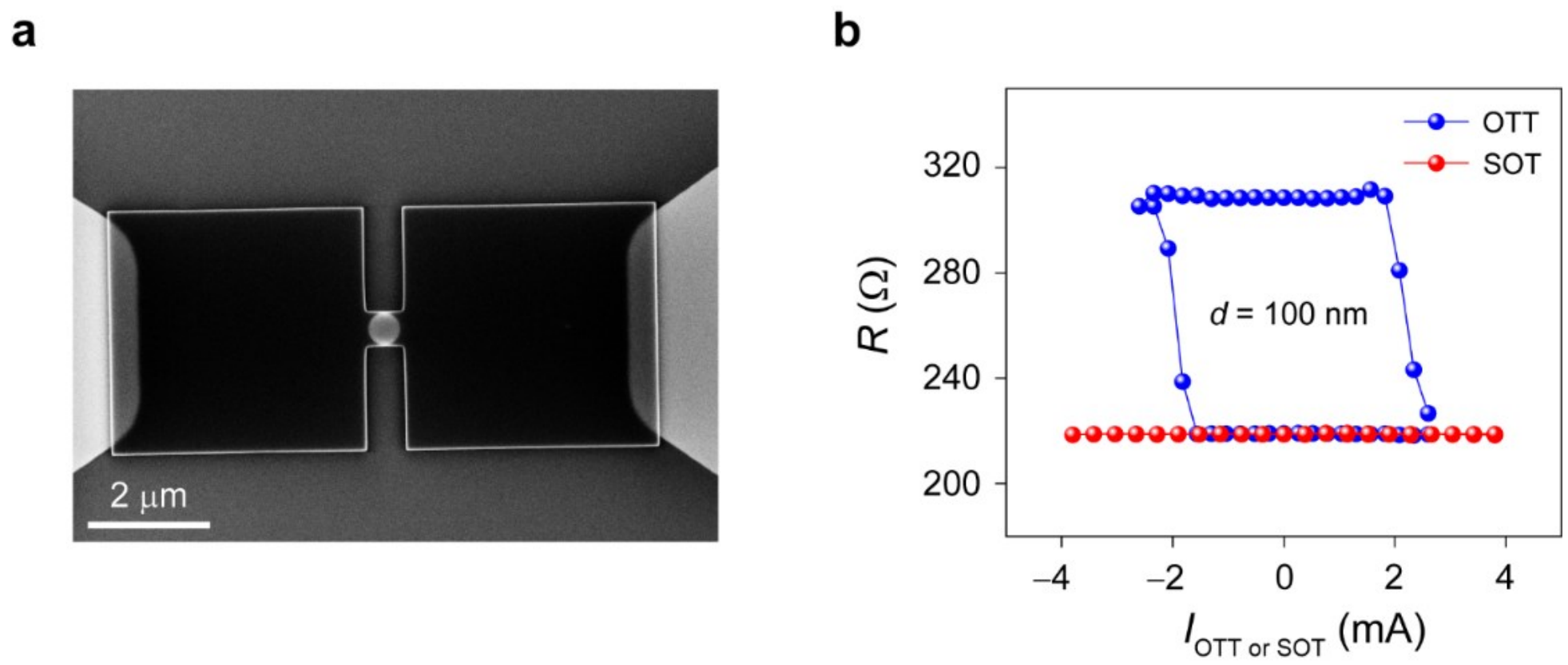

**Figure S6| Comparison of two- and three-terminal switching measurements in AATJs. a,** A representative SEM image of the AATJ devices used for this experiment, prior to the top electrode deposition. The AATJ pillar is located at the center of the narrow Pt channel whose width is slightly larger than the diameter of the pillar. **b**, Three-terminal SOT and two-terminal OTT switching measurements performed within the same device with a 100 nm pillar diameter. The blue and red curves indicate currents applied vertically through the AATJ, and horizontally through the underlying Pt channel, respectively. The absence of switching, even for in-plane currents that are ~2x larger than the observed OTT switching current, confirms that SOT is not the primary mechanism of the observed two-terminal switching signatures.

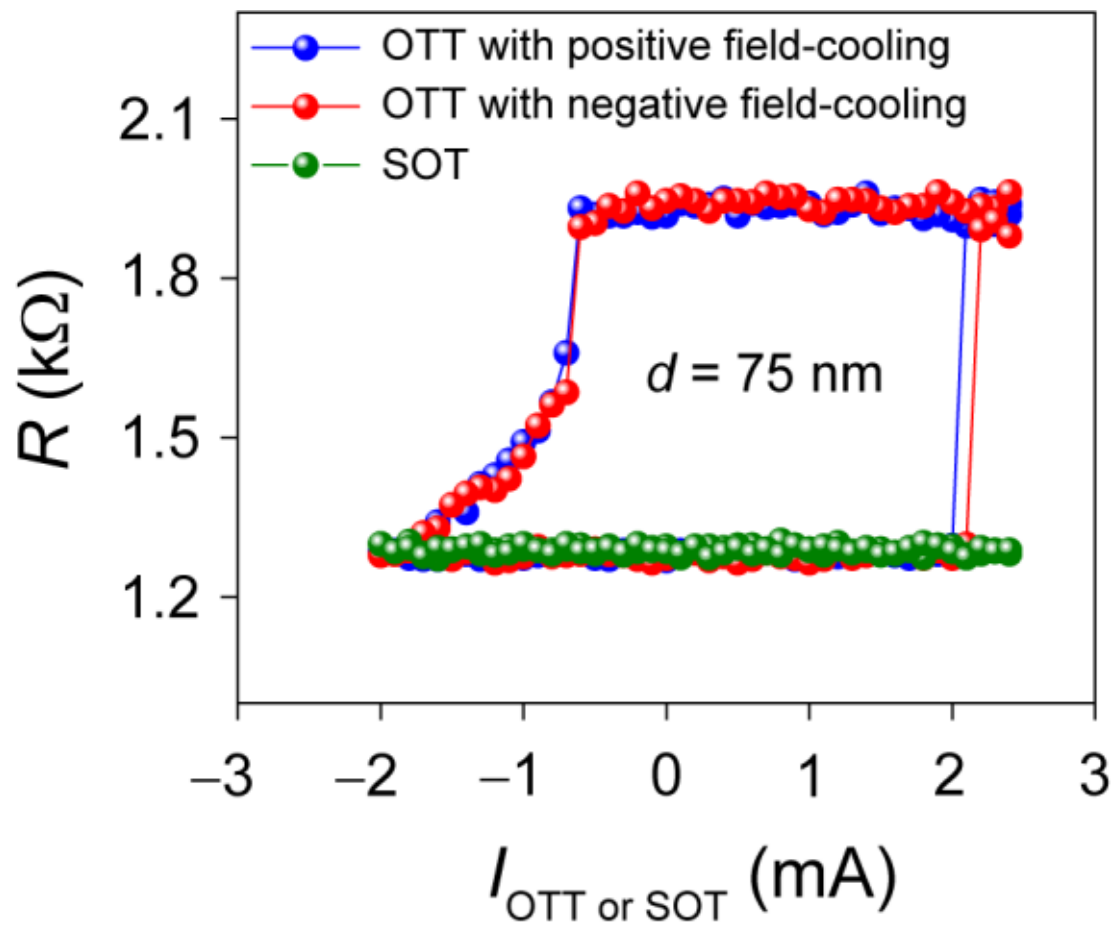

**Figure S7| Comparison of two- and three-terminal switching measurements of an AATJ under different field-cooling conditions.** Three-terminal SOT and two-terminal OTT switching

measurements performed within the same device with a 75 nm pillar diameter. Prior to the OTT switching, the AATJ was field-cooled from 400 K to 300 K under an out-of-plane external magnetic field of ±0.2 T, respectively. The blue and red curves indicate write currents applied vertically through the AATJ (i.e., OTT geometry) while the green curve indicates write currents applied horizontally through the underlying Pt channel (i.e., SOT geometry). All measurements were conducted at room temperature.

**Supplementary Note 6. Micromagnetic simulations of switching in the presence of domains**

To account for domain nucleation, domain wall motion, and their size dependence in AATJs, we developed a micromagnetic model while retaining the same magnetic parameters as those used for the macrospin model of the main text, which was built upon our earlier-developed two-sublattice model [3]. Within the micromagnetic approach, the antiferromagnet is discretized in cuboid cells, each containing three sublattice magnetizations. The magnetization dynamics for each cell are given by three coupled LLG equations, one per each sublattice ($\vec{m}_1$, $\vec{m}_2$, $\vec{m}_3$), as described by Eq. (9) in the main text. Formally, this model is the same as the macrospin model. However, here the effective field, $H_{\text{eff}}$, includes the exchange contribution arising from the coupling between the same sublattices in neighboring cells. This term, equivalent to ferromagnetic exchange, promotes parallel alignment of the spins of one sublattice in different points and is expressed as $\vec{H}^0_{\text{exch}} = 2A_0\nabla\vec{m}_i/M_{s,i}$. We have considered each sublattice equivalent and characterized by the exchange constant $A_0 = 5$ pJ/m, even though we verified that there are no significant changes in the results for a wide range of values from 2.5 to 10 pJ/m.

We studied the OTT dynamics for three pillar diameters (50, 100, 200 nm) at a temperature of 300 K. We modeled the free layer directly and considered the effect of the polarizer layer (i.e., fixed $PtMn_3$ electrode) via the corresponding sublattice polarizations entering the OTT equations, as presented in Eq. (9) of the main text. For each diameter, we first relaxed the magnetic configuration starting from a head-to-head domain configuration along each sublattice anisotropy direction of the Kagome arrangement. We found that for small diameters (50 nm) the magnetic texture remains uniform, while for 100 and 200 nm diameters, domains are nucleated. These are shown in Fig. S8a and Fig. S8b for the diameters of 100 and 200 nm, respectively.

We then investigated the OTT dynamics by applying a 50 ps square-shaped current pulse, and letting the system relax for another 10 ps, in the following three cases:

I) Both the polarizer and free layer are uniformly magnetized.

II) The polarizer is uniform, while the relaxed non-uniform states are used for the free layer.

III) The free layer is uniform, while the relaxed non-uniform states are used for the polarizer.

*Case I.* The switching is always coherent, independent of the pillar diameter. This case is adequately captured by a macrospin model.

*Case II.* The switching always yields a uniform final state within the pulse duration (50 ps). This is shown in Fig. S8c and Fig. S8d, where we present the time-dependent $x$, $y$ and $z$ components of the sublattice magnetization $\vec{m}_3$ for 100 and 200 nm pillars, respectively, and in Fig. S8e and Fig. S8f, where the snapshots at different times are depicted. The magnetizations of the other two sub-lattices are rotated by 120° in the plane. There is an initial domain wall propagation in response to the current, which ultimately leads to a uniform state in the free layer, following the uniform configuration of the polarizer.

*Case III.* The non-uniform magnetic configurations, depicted in Fig. S9a and Fig. S9b, are now utilized to describe the non-uniform sublattice-resolved spin polarization that exhibits a strong dependence on the pillar diameter. The free layer magnetization, however, is assumed to be initially uniform. For the 100 nm case, the free layer magnetization evolves via coherent rotation, as shown in Fig. S9c and Fig. S9e. Once the current is turned off, the magnetization of each sublattice relaxes towards the equilibrium in-plane Kagome configuration while staying relatively uniform.

For the larger 200 nm diameter case, the dynamics are more complex, as shown in Fig.

S9d. In Fig. S9b, we can see regions with different in-plane and out-of-plane directions of the polarization for each sublattice. Hence, there will be regions where the free layer anisotropy directions for each sublattice and the respective sublattice-dependent spin polarization direction are aligned, and others where they are misaligned. Therefore, the current pulse results in the formation of domains, and domain wall-mediated dynamics associated with a rotation of the magnetic texture induced by the OTT, as observed in Fig. S9f. In this case, we can expect a reduction in TMR due to the partial misalignment of the free layer sublattice magnetizations with respect to those of the polarizer.

Therefore, our results indicate that in general, AATJs with larger pillar diameters will be characterized by non-uniform magnetic textures and non-uniform magnetization dynamics, which can reduce the effective TMR compared to AATJs with uniformly magnetized free and fixed layers.

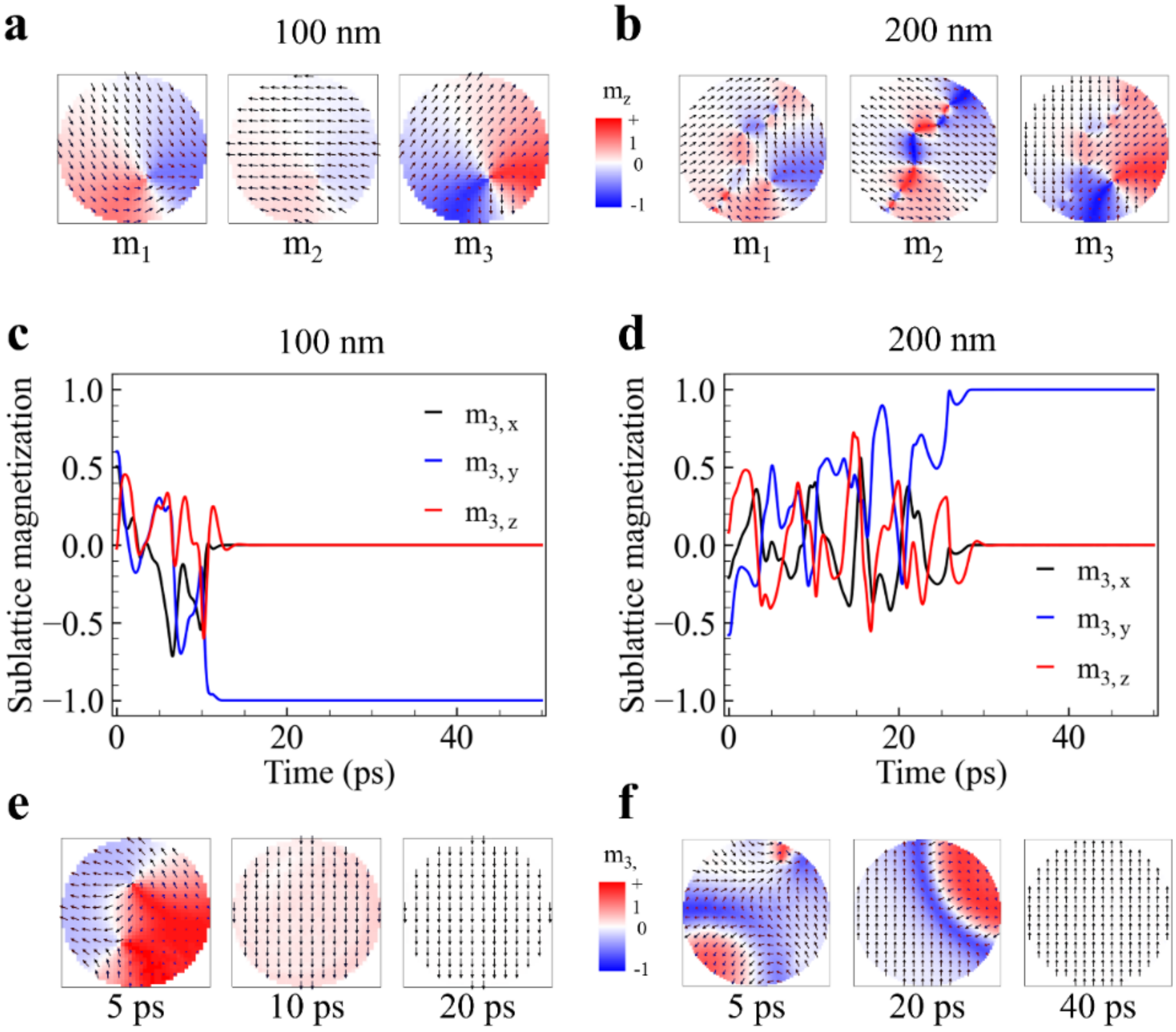

**Figure S8| Micromagnetic simulations of the size dependence of OTT dynamics in the case of non-uniform free layer magnetization. a, b,** Snapshot of the equilibrium magnetic configuration of the free layer for pillars with diameter of 100 nm **(a)**, and 200 nm **(b)**. **c, d,** Time-dependent *x*,*y* and *z* components of the magnetization of sublattice 3 in response to a 50 ps current pulse for pillars with diameter of 100 nm **(c)**, and 200 nm **(d)**. **e, f,** Snapshots taken at different times during the switching dynamics of sublattice 3 for pillars with diameter of 100 nm **(e)**, and 200 nm **(f)**.

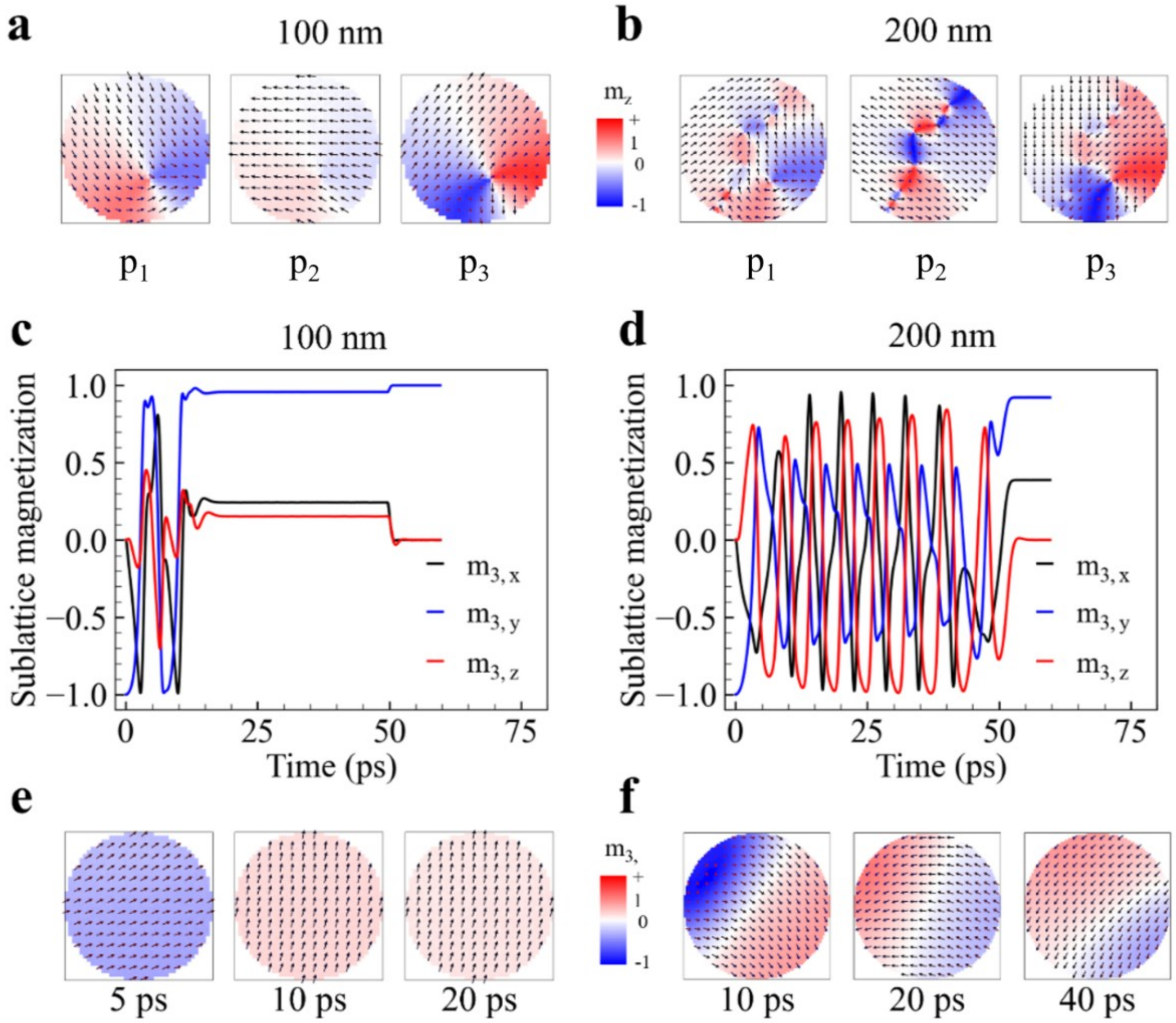

**Figure S9| Micromagnetic simulations of the size dependence of OTT dynamics in the case of non-uniform fixed layer polarization. a, b,** Snapshot of the current-induced polarization directions used for pillars of diameter 100 nm **(a)**, and 200 nm **(b)**. **c, d,** Here, we have used the same relaxed states as in Fig. S8, but assumed them to represent the equilibrium state of the polarizer. Time-dependent *x*,*y* and *z* components of the magnetization of sublattice 3 of the free layer in response to a 50 ps current pulse for pillars with diameter of 100 nm **(c)**, and 200 nm **(d)**. **e, f,** Snapshots taken at different times during the switching dynamics of sublattice 3 of the free layer for pillars with diameter of 100 nm **(e)**, and 200 nm **(f)**. For all the snapshots the palette describes the magnitude of the z-component and the arrows represent the in-plane components.

**Supplementary Note 7. Anomalous Hall effect measurements**

To investigate the magnetic-field-induced transport response, we performed anomalous Hall effect (AHE) measurements on two sets of devices, the results of which are shown in Fig. S10:

1. **Poly-crystalline $PtMn_3$:** Full material stacks of Si / $SiO_2$ / Pt (5 nm) / $PtMn_3$ (7 nm) / MgO (2 nm) / $PtMn_3$ (10 nm) / Pt (5 nm) / Ru (10 nm), nominally identical to the AATJ material stacks of the main manuscript, were patterned into Hall-bar structures with a width of 10 μm. We did not observe an AHE signal in these samples, as shown by the blue data points in Fig. S10b. This result is consistent with the absence of a net magnetic moment in this stack, as confirmed by SQUID magnetometry measurements, indicated by blue data points in Fig. S10a.

2. **Epitaxial $PtMn_3$:** We deposited epitaxial $PtMn_3$-based stacks of Pt (7 nm) / $PtMn_3$ (30 nm) on an $Al_2O_3$ (0001) substrate. The $PtMn_3$ layer was capped by a layer of $AlO_x$ (2 nm) for protection. The Pt layer was grown at the substrate temperature of 500˚C, followed by post-annealing at 850˚C for one hour to promote epitaxial growth. Subsequently, $PtMn_3$ was deposited at the substrate temperature of 550˚C. To examine the crystallinity of the material stacks, we performed X-ray diffraction (XRD) as shown in Fig. S10c. The XRD measurement shows a clear peak at $2\theta = 40.2^\circ$ whose asymmetric shape indicates the overlap of the Pt (111) and $PtMn_3$ (111) peaks. The red data points in Fig. S10b show the AHE signal for this sample. The presence of AHE is consistent with the presence of a small magnetic moment, as confirmed by SQUID measurements, indicated by red data points in Fig. S10a.

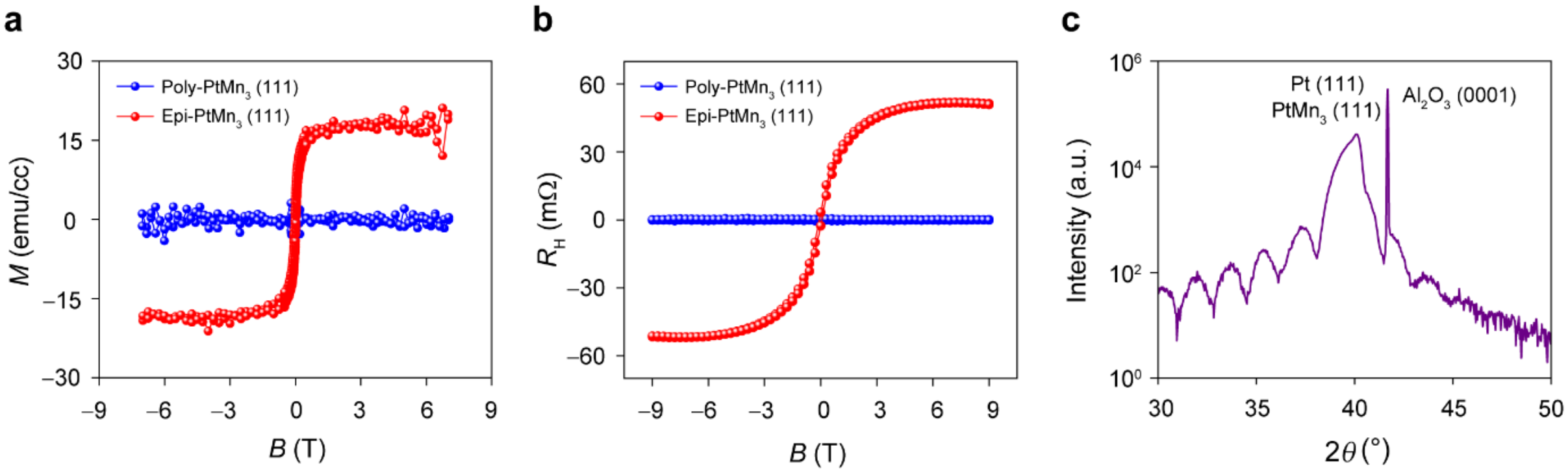

**Figure S10| Anomalous Hall measurements in poly-crystalline and epitaxial $PtMn_3$ (111) films. a,** SQUID magnetometry results on poly-crystalline (blue) and epitaxial (red) $PtMn_3$ films. **b,** AHE measurements on Hall bars made from the same samples, indicating a non-zero response in the epitaxial $PtMn_3$ case. **c,** XRD measurements on the epitaxial $PtMn_3$ sample, indicating the $PtMn_3$ (111) peak as well as the $Al_2O_3$ (0001) peak from the substrate.

**Supplementary Note 8. Two-terminal current-induced switching under external magnetic fields**

To demonstrate the immunity of the OTT-controlled AATJs to external fields, we performed two-terminal current-induced switching measurements under three different out-of-plane externally applied magnetic fields ($B_z$). Figure S11 presents the switching results when $B_z = +2$, 0 and −2 T. All three curves exhibit similar switching behavior without any significant changes in switching current values or TMR, indicating the resilience of the OTT switching against external magnetic fields.

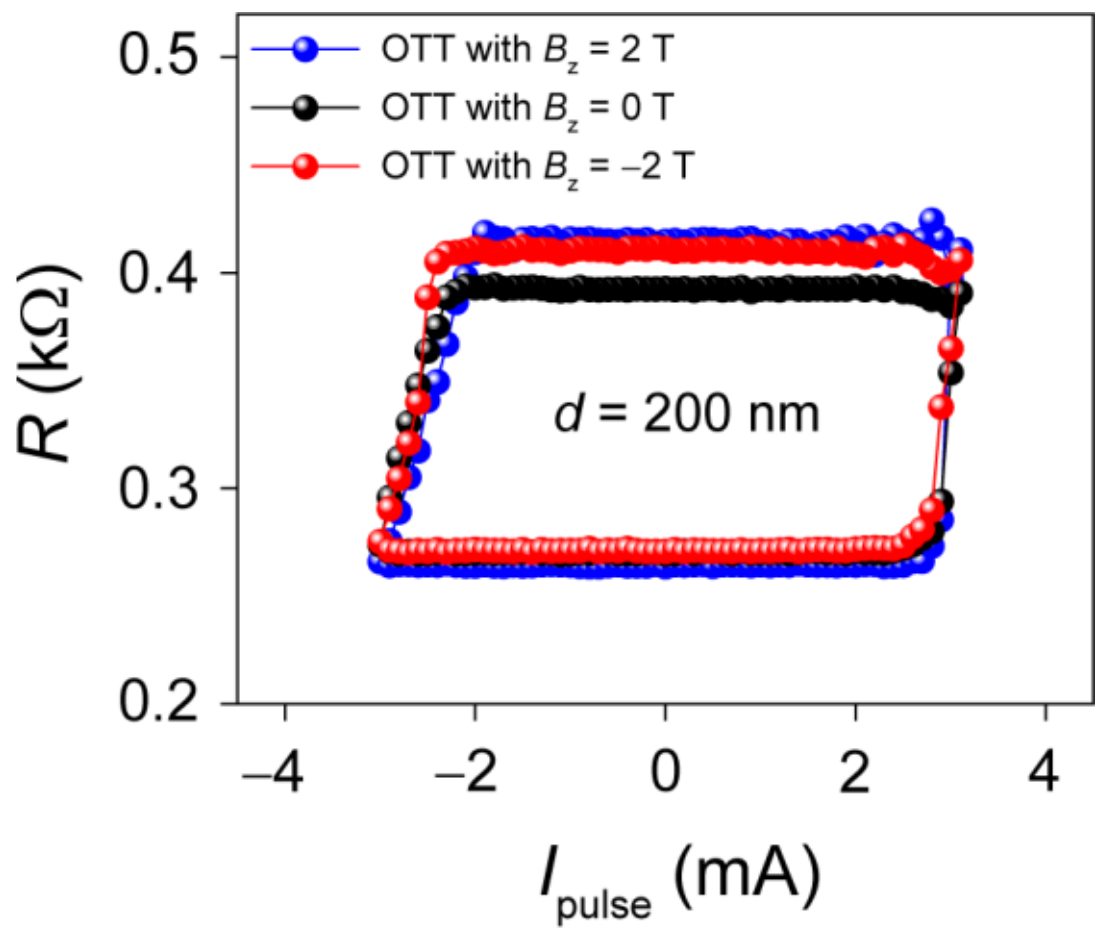

**Figure S11| Two-terminal current-induced switching measurements of an AATJ under different magnetic fields.** Out-of-plane external magnetic fields of +2, 0, −2 T were applied during the switching measurements on a 200 nm-diameter AATJ at room temperature, corresponding to the blue, black and red curves, respectively.

## References

[1] Körner, W., Gumbsch, P., & Elsässer, C. Analysis of electronic subgap states in amorphous semiconductor oxides based on the example of Zn-Sn-O systems. *Phys. Rev. B* **86**, 165210 (2012).

[2] Grimaldi, E., Krizakova, V., Sala, G., Yasin, F., Couet, S., Kar, G. S., Garello, K., & Gambardella, P. Single-shot dynamics of spin–orbit torque and spin transfer torque switching in three-terminal magnetic tunnel junctions. *Nat. Nanotechnol.* **15**, 111-117 (2020).

[3] Puliafito, V., Khymyn, R., Carpentieri, M., Azzerboni, B., Tiberkevich, V., Slavin, A., & Finocchio, G. Micromagnetic modeling of terahertz oscillations in an antiferromagnetic material driven by the spin Hall effect. *Phys. Rev. B* **99**, 024405 (2019).